\documentclass[conference]{IEEEtran}
\IEEEoverridecommandlockouts
\input{preamble.sty}
\newcommand{\setOfReals}{\mathbb{R}}
\newcommand{\setOfNaturals}{\mathbb{N}}

\newcommand{\indep}{\perp \!\!\! \perp}

\newtheorem{theorem}{Theorem}[section]

\newtheorem{lemma}[theorem]{Lemma}
\newtheorem{fact}[theorem]{Fact}
\newtheorem{propo}[theorem]{Proposition}

\makeatletter
\newcommand{\distas}[1]{\mathbin{\overset{#1}{\kern\z@\sim}}}%
\newsavebox{\mybox}\newsavebox{\mysim}
\newcommand{\distras}[1]{%
  \savebox{\mybox}{\hbox{\kern3pt$\scriptstyle#1$\kern3pt}}%
  \savebox{\mysim}{\hbox{$\sim$}}%
  \mathbin{\overset{#1}{\kern\z@\resizebox{\wd\mybox}{\ht\mysim}{$\sim$}}}%
}
\makeatother

\newboolean{todo}
\setboolean{todo}{true} 
\newcommand{\remarkInternal}[4]{\ifthenelse{\boolean{todo}}{\todo[inline, color=#2, caption={2do}, #3]{\begin{minipage}{\textwidth-4pt}\emph{Remark #1:}\\#4\end{minipage}}}{}}

\usepackage{balance}

\usepackage{mathtools}

\usepackage{stfloats}

\def\BibTeX{{\rm B\kern-.05em{\sc i\kern-.025em b}\kern-.08em
		T\kern-.1667em\lower.7ex\hbox{E}\kern-.125emX}}
\begin{document}
	
	\title{Optimal Decision Making in Active Queue Management}
	
	
\author{

\IEEEauthorblockN{%
    Sounak Kar\IEEEauthorrefmark{1},
    Bastian Alt\IEEEauthorrefmark{2},
    Heinz Koeppl\IEEEauthorrefmark{2} and
    Amr Rizk\IEEEauthorrefmark{3}%
  }%
  \IEEEauthorblockA{\IEEEauthorrefmark{1} EPFL, {\tt\small sounak.kar@epfl.ch}}%
  \IEEEauthorblockA{\IEEEauthorrefmark{2} Technical University of Darmstadt, {\tt\small firstname.lastname@bcs.tu-darmstadt.de}}%
  \IEEEauthorblockA{\IEEEauthorrefmark{3} University of Duisburg-Essen, {\tt\small amr.rizk@uni-due.de}}%
}

\maketitle

\begin{abstract}

Active Queue Management (AQM) aims to prevent bufferbloat and serial drops in router and switch FIFO packet buffers that usually employ drop-tail queueing. AQM describes methods to send \textit{proactive} feedback to TCP flow sources to regulate their rate using selective packet drops or markings. Traditionally, AQM policies relied on heuristics to approximately provide Quality of Service (QoS) such as a target delay for a given flow. These heuristics are usually based on simple network and TCP control models together with the monitored buffer filling. 
A primary drawback of these heuristics is that their way of accounting flow characteristics into the feedback mechanism and the corresponding effect on the state of congestion are not well understood.
In this work, we show that taking a probabilistic model for the flow rates and the dequeueing pattern, a Semi-Markov Decision Process (SMDP) can be formulated to obtain an optimal packet dropping policy. 
This policy-based AQM, named \textit{PAQMAN}, takes into account a steady-state model of TCP and a target delay for the flows. Additionally, we present an inference algorithm that builds on TCP congestion control in order to calibrate the model parameters governing underlying network conditions. 
Using simulation, we show that the prescribed AQM yields comparable throughput to state-of-the-art AQM algorithms while reducing delays significantly.

\textbf{Keywords}: Active Queue Management, Markov Decision Processes.
\end{abstract}

\section{Introduction} \label{sec:intro}

Striking a balance between the two most common metrics of performance in IP networks, i.e., throughput (or utilization) and delay is a fundamental problem for routing devices.
Active queue management (AQM) has evolved as a mechanism to augment prevalent end-system protocols such as TCP to tune the performance in terms of the said metrics.
Most modern network switches have built-in buffers, which accumulate incoming data packets while the switch is busy processing and transmitting processed packets to the output ports~\cite{iyer2002routers}. 
A large buffer has the advantage that it can potentially minimize droptail packet loss, i.e., dropping incoming packets when the buffer is full. 
Understandably, this leads to higher network utilization while causing excess buffering of packets and, consequently, longer delays. 
This phenomenon is commonly referred to as bufferbloat~\cite{gettys2011bufferbloat}.
Using a shallow buffer alleviates this problem, although at the expense of frequent packet drops, which may lead to diminished throughput of TCP flows\footnote{In this work, we consider only TCP flows as they constitute the vast majority of the internet traffic according to estimates such as~\cite{5476872}.}~\cite{chiu1989analysis}. 
Evidently, the trade-off between higher utilization and lower queueing delay is a matter of policy~\cite{floyd2001adaptive}, which is delineated by the implemented AQM algorithm at the switch. 
The drop/admit decision usually depends on inputs such as current buffer-filling, packet delay history, or recent packet drop pattern. \looseness=-1

In the last three decades, a range of algorithms have been proposed~\cite{redAqm, codelAqm, pieAqm} to address the \emph{AQM problem}, i.e., achieving an acceptable trade-off between link utilization and packet delay. 
Occasionally, these algorithms focus on additional aspects such as scalability, robustness, or fairness; for an extensive survey,
see~\cite{aqmSurvey} and the references therein. 
The first AQM algorithm RED~\cite{redAqm}, proposed by Floyd and Jacobson in 1993, calculates an exponentially weighted average queue length and, as a linear function of this average, computes an initial packet drop probability in $[0,p_\text{max}]$.
To avoid serial drops, the initial probability is further transformed into a final drop probability that takes into the account the number of packets admitted since the last packet drop. 
An incoming packet is dropped according to this probability if the average queue length is between two threshold parameters for queue length and dropped/admitted deterministically otherwise.
Clearly, the resulting policy depends critically on these threshold parameters and $p_\text{max}$. 
Although RED is shown to be fairer to bursty traffic than the classic drop tail~\cite{redAqm}, \textit{the main challenge lies in identifying the model parameters and there has been no consensus on the corresponding parameter engineering process.}
To get around this problem, a range of variants and extensions of RED have been proposed, which address this issue with fairly limited success. 
We again refer the reader to~\cite{aqmSurvey} for a thorough discussion on these algorithms.\looseness = -1

A popular AQM algorithm that claims to have successfully circumvented the problem of parameter engineering is CoDel~\cite{codelAqm}. Although CoDel is driven by two input parameters, one that signifies a \textit{target delay} (at the AQM enabled node) and a window parameter specifying how often a packet should be dropped, their default value is hardly changed in practice.
Starting with the default value of the window parameter, CoDel successively adapts its value until it meets the target delay. 
Although claimed to be knob-free due to the prescribed default values of the parameters under universal traffic conditions, these values can be tuned to yield superior performance for a given environment~\cite{khademi2014new}. 
However, the analysis for the choice of the default values under varied conditions has been limited to minimal empirical investigations~\cite{khademi2014new}. \looseness = -1

Different implementations exist for CoDel: although originally devised to drop a packet after it is already enqueued, it can also be implemented to obtain the current queueing delay and drop packets at the ingress depending on the information exposed on the data plane\footnote{This makes CoDel simpler to implement on modern programmable data plane devices~\cite{Kundel21}.}.
Similarly, the AQM algorithm called PIE~\cite{pieAqm} drops packets directly at the input port. 
PIE calculates the drop probability of an incoming packet by looking at the current queue-filling and the departure rate from the queue. 
This further improves the processing overhead compared to CoDel, which requires calculation of delay per packet. 
Additionally, the consideration of the current queue length to calculate the drop probability implies that congestion is directly controlled. 
Although claimed to be knob-free like CoDel, PIE actually requires default values for target delay, drop frequency and parameters for drop probability adaptation~\cite{khademi2014new}. 
Further, the drop probability parameters are adapted according to a rule that is based on judgements. 
This phenomenon extends to most, if not all, popular AQM algorithms of today~\cite{khademi2014new}.

In light of the above, we propose a Markov Decision Process (MDP)-based approach to address the AQM problem, called PAQMAN, which only requires the target delay and the relative importance of delay violation to throughput reduction as inputs. 
To detect potential congestion, PAQMAN uses the current queue-filling, an estimate of the arrival rate as well as estimates of the flow RTTs.
We make the case that to optimally decide on packet drops, it is necessary for an AQM algorithm to account for such a holistic description of the system state.
In our framework, we encode the optimization goal through a reward function that combines delay and throughput objectives and reflects the immediate gain following a packet drop/admit decision. 
Consequently, we derive an optimal policy (the AQM algorithm) using tools from the MDP framework that take the state transition probabilities calculated from the model and the reward function as inputs.
The policy provides the optimal decision for every possible instance of the system state.
For the simpler case where the switch deals with a single flow having negligible RTT, we model the packet arrival process in greater generality, whereas a more tractable model is chosen for the general case with multiple flows.
Finally, we compare the performance of PAQMAN with the state-of-the-art algorithm CoDel and the classic droptail queues. 
Our findings show that PAQMAN yields equivalent utilization/throughput to CoDel while minimizing delays considerably. 

It should however be noted that PAQMAN, like most algorithms, is not universally applicable and is rather based on specific assumptions that are not always met in practice.
This leads to following \emph{limitations} for PAQMAN:
\begin{itemize}
    \item The analytical underpinning of PAQMAN is  derived based on TCP flows using the \emph{same} congestion control algorithm.
    \item For TCP flows with non-negligible RTT, PAQMAN requires an estimate of flow RTT, which can be obtained by passive measurement in certain cases~\cite{jiang2002passive,veal2005new,arouche2014machine} with considerable accuracy. 
    \item While dealing with flows with non-negligible RTT, the analytical derivation of PAQMAN also assumes that the packet interarrival times can be modelled by exponential distribution. 
    \item PAQMAN is based on a MDP model for which solutions suffer from the curse of dimensionality. Thus, like many AQM algorithms, the implementation of PAQMAN for many concurrent flows will impose memory requirement. A lighter version that trades accuracy for ease of implementation, e.g., through neural networks, is left as future work.
    \item For the multiple flows with non-negligible RTTs, the current version of PAQMAN is trained offline and implemented later. 
    \item We do not investigate the aspect of fairness for PAQMAN and leave it for future work.
\end{itemize}

The remainder of the paper is structured as follows: we first an overview of the related work in Sect.~\ref{sec:relatedWork} and subsequently formulate the AQM decision problem in Sect.~\ref{sec:model}. 
Sect.~\ref{sec:model} describes the algorithm when the switch deals with a single flow having negligible RTT, while we discuss the general case with non-negligible RTTs in Sect.~\ref{sec:rttModel}. 
Finally, we present numerical simulation results in Sect.~\ref{sec:numerical} and conclude the paper in Sect.~\ref{sec:conclusion}.
\section{Related Work} \label{sec:relatedWork}
AQM was developed as an additional module on top of congestion control functions with the aim to keep the congestion levels lower than traditional droptail queues.
This was achieved by sending early congestion signals based on congestion indicator(s), which might include queue length, packet arrival/departure rate, flow round-trip time, link capacity, and number of flows.
While the congestion signal was traditionally in form of a packet drop, ECN markings also came into use to enhance throughput. 
Further, the packet drop action for an AQM can be random as against deterministic packet drops in droptail queues.
In this section, we compare these aspects of PAQMAN to that of the first AQM algorithm RED\cite{redAqm}, as it formed the basis of AQM research, and the two most popular AQM as of today: PIE~\cite{pieAqm} and CoDel~\cite{codelAqm}.
Occasionally, we also refer to variants of these AQMs.

As for congestion indicator, RED uses a moving average of queue length.
Depending on a higher and a lower threshold parameter for queue length, an incoming packet is dropped or admitted if the (calculated) average queue length is outside these threshold, while, within these thresholds, the packet is dropped according to a drop probability that is a linear function of the average queue length.
Similarly, PIE looks at the queue length and departure rate from the queue to calculate a drop probability, while CoDel uses the minimum delay of dequeued packets over an observation window as its congestion indicator.
In comparison, PAQMAN uses the current packet arrival rate and the queue length to detect early congestion in the canonical case with one flow having negligible RTT. 
For the general case with multiple flows having non-negligible RTT, the flow index of the incoming packet, the time since the last decision event, and the RTTs are additionally included in the set of congestion indicators. 
Thus, we have a more holistic description of the state of the switch under PAQMAN, allowing us to account for more uncertainties impacting the state evolution. 

\begin{figure*}[t]
    \centering
    \includegraphics[width=0.8\textwidth]{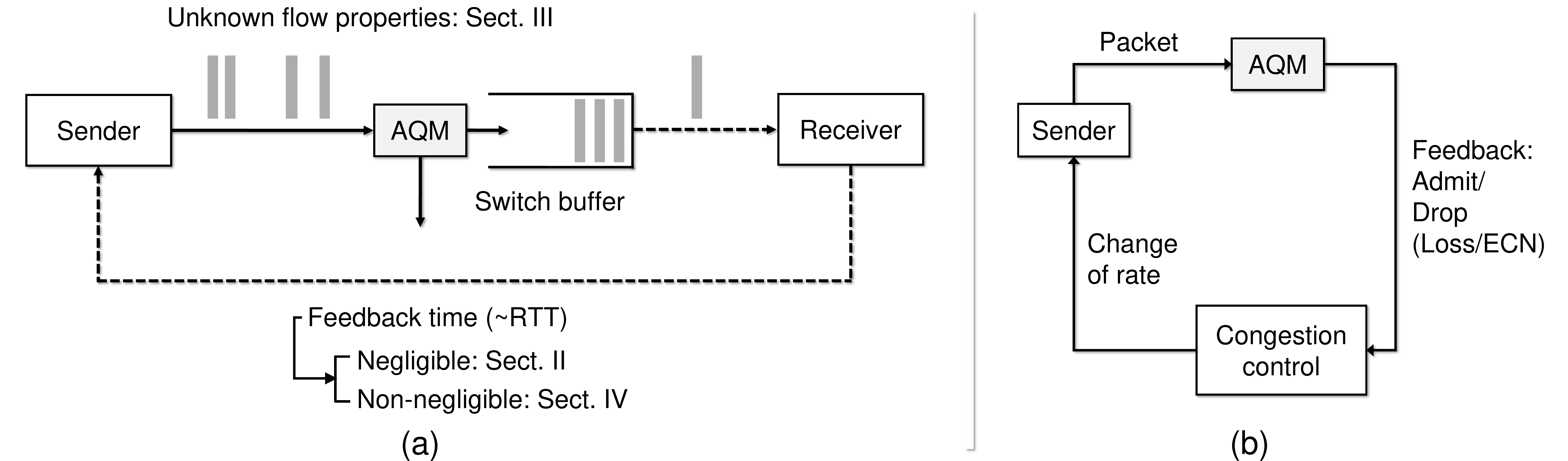}
    \caption{ Part (a) shows how PAQMAN distinguishes between different AQM scenarios: the time until the receipt of the feedback of drop/admit action is the transmission time between the switch and the sender via the receiver, which is approximated as one RTT. The case with negligible feedback time (RTT) is discussed in Sect.~\ref{sec:model} and the other case is taken up in Sect.~\ref{sec:rttModel}. When traffic properties are unknown, we estimate corresponding parameters according to Sect.~\ref{sec:inference}. Part (b) highlights how AQM acts in tandem with the underlying congestion control by sending early congestion signals and thereby influencing the sending rate to achieve a desired level of congestion.
    }
    \label{fig:dataCentre}
    \vspace{-10pt}
\end{figure*}

In terms of packet drop action, RED can either drop randomly (according to a certain probability) or deterministically (i.e., drop or admit), depending on the average queue length.
While CoDel uses deterministic drop action, PIE always calculates a drop probability to implement its random drop action.
Similar to CoDel, PAQMAN's drop action is deterministic.
In terms of congestion signal, packets are dropped under RED whereas they are ECN-marked in a switch deploying REM~\cite{lapsley1999random}. 
While we use packet drop for PAQMAN throughout this paper, it can also be used in conjunction with ECN marking and the corresponding changes in the derivations are straightforward. We highlight this in part (b) of Fig.~\ref{fig:dataCentre}, where the AQM is seen as an intermediate module between the sender and the congestion control mechanism of the network.

The core of the AQM algorithm can be viewed as a rule which translates the congestion indicator into the drop/admit action or, more generally, the likelihood of the same. 
There are primarily three approaches~\cite{aqmSurvey} to formulate this rule: heuristic-based, control theoretic, and deterministic optimization-based. 
While the early algorithms such as RED, its variants~\cite{floyd2001adaptive,ott1999sred} and the popular algorithm CoDel follow heuristic approaches, the newer algorithms (such as PIE) adopted control theoretic framework to circumvent the judgemental aspects of the former approaches such as parameter tuning. 
For simplicity and tractability, most of the control theoretic algorithms build on the Additive Increase Multiplicative Decrease (AIMD) principle of TCP ignoring other aspects like slow start and retransmission timeouts. 
Further, these works assume a linearized fluid model of TCP proposed in~\cite{hollot2001control}. 
Some prominent examples under the control theoretic approach include: Proportional-Integral~\cite{hollot2001control}, Proportional-Derivative~\cite{kim2003design}, Proportional-Integral-Derivative~\cite{yanfie2003design}, and Proportional-Integral-Enhanced~\cite{pieAqm} controller. 
For a thorough comparison of the algorithms under this approach, see Table VI-IX of~\cite{aqmSurvey}.
Recently, an information compression approach has been adopted in~\cite{liu2018adaptive}, where starting from the TCP fluid model, a simpler relation between queueing delay and drop probability is formulated. 
The control law here specifies the change in drop probability to drive the delay towards a reference value. 
We note that the control theoretic approaches in general require certain input parameters whose default values are in the end set according to judgement. 
In comparison, PAQMAN employs an MDP-based approach and only requires the target delay and the relative importance of target delay violation to reduction in throughput as input parameters. 

Works that adopt MDP as their choice of tool include~\cite{alpcan2006control,alpcan2008non}, where the authors first assume that the system dynamics is given by a deterministic fluid model for which they set out to perform a non-equilibrium analysis. 
Here, they first identify a Markov chain that closely approximates the behaviour of the deterministic system, i.e., estimate the transition matrix of the chain for a given level of state-space discretization.
Subsequently, they adopt an MDP-based approach to derive the best policy from a set of candidate policies comprising drop-tail, RED, or a interpolated version of these two. 
In comparison, we use the MDP framework to derive the policy itself. \looseness = -1

The MDP framework has been used directly for congestion control as well.
While our objective is to determine a packet admission rule, the paper~\cite{wang2021intelligent} focuses on update of congestion window using DQN.
We see that most literature in this space uses closely-related but a different (non-binary) action space.
For example,~\cite{wu2002burst} sets the transmission rates of constituent flows in closed intervals for the next epoch, while~\cite{habachi2013online} updates the congestion window for multimedia content using a POMDP framework.
Optimal drop/admit actions were considered in~\cite{jacko2012optimal} in the context of anticipative congestion control in a network of routers, where the objective is not to waste resources on processing a packet if it is likely to be dropped downstream. \looseness = -1



In addition to delay-utilization trade-off, the aspects of drop rate, jitter and fairness also gained focus in AQM literature over the years. 
However, there is no standardized evaluation criteria for AQM schemes and the authors in~\cite{bitorika2003evaluation} attribute the slow progress of AQM research to this fact. 
In addition to the measures above, the authors in~\cite{aqmSurvey} consider scalability, stability, responsiveness, and robustness to be crucial metrics of performance. 
Scalability focuses on the feasibility of implementation of an AQM algorithm as the number of flows increases, whereas stability measures, for example, the change in queue length as the number of flows varies.
Further, the speed of convergence is referred to as responsiveness~\cite{hollot2003unresponsive} and robustness signifies the ability of an AQM algorithm to work under diverse network conditions. 
Thus, robustness necessitates dynamic parameter tuning to suit changing traffic loads, i.e., variation of network parameters.

Following the inclusion of varied performance measures, RED was modified to meet the new objectives. 
For example, authors in~\cite{ott1999sred} introduced SRED to enhance stability, ARED~\cite{floyd2001adaptive} used auto-tuning to increase robustness, and FRED~\cite{lin1997dynamics} focused on fairness. 
We refer the reader to Table XIV- XXIV of~\cite{aqmSurvey} for a detailed comparison of the heuristic AQM schemes that address certain aspects of performance.
Coming back to the issue of dynamic parameter tuning, the paper~\cite{gomez2019intelligent} introduces a general parameter tuning method that can be used in conjunction with any given AQM algorithm. 
Independently of the underlying AQM algorithm, the problem of parameter tuning is formulated in~\cite{gomez2019intelligent} as an MDP, where the (discretized) predicted value of the congestion indicator denotes the state of the system and the action set consists of all possible values of the AQM parameter. 
Taking the measured throughput-to-RTT ratio as the immediate reward, the authors adopt a Q-learning approach~\cite{watkins1992q} to find the optimal AQM parameter value for a given estimate of the congestion predictor in the next time interval. \looseness = -1

As mentioned, delay-throughput trade-off   has been the primary objective of AQM algorithms. 
Later, most control theoretic approaches additionally focused on stability, responsiveness, and robustness by analyzing the transient response, the oscillation around the target queue length, and the steady-state error, respectively. 
However, fairness is usually not measured under control theoretic as well as deterministic optimization approaches. 
In comparison, we empirically evaluate PAQMAN in terms of delay, throughput, 
and the rate of convergence to steady-state behaviour. However, we do not evaluate its performance in terms of jitter, responsiveness, scalability, or robustness. [Todo: Fairness].

The algorithms that dominate the present-day AQM landscape are PIE and CoDel. They have been compared extensively to each other and to some variants of RED~\cite{jarvinen2014evaluating,khademi2014new}. 
The former paper considers both performance and scalability under multiplexing. 
It concludes that CoDel visibly leads to better performance in terms of delays, while the performance of PIE scales well for multiplexed flows. 
Similarly, the authors in~\cite{khademi2014new} recommend CoDel over PIE after their extensive evaluation over the range of respective default parameters, as the empirical delay distribution under PIE was seen to have a longer tail. 
Hence, in this work, we evaluate PAQMAN against CoDel and droptail as they can be justly considered as AQM and non-AQM benchmarks, respectively.

\vspace{4pt}
\makebox[0.95\columnwidth]{\fbox{
\begin{minipage}{0.95\columnwidth}
  {\vspace{2pt}
  
  \noindent $Q_t$: queue length immediately before $t$-th packet arrival, \newline
  $\alpha$: common Gamma shape parameter of packet interarrival time distributions, \newline
  $\beta_t$: gamma rate parameter of the packet interarrival time immediately before $t$-th arrival, \newline
  $S_t$: state of the system observed by the packet corresponding to $t$-th arrival, given by $(Q_t, \beta_t)$, \newline
  $\mu$: service rate of the packets, \newline
  $A_t$: action taken upon $t$-th arrival, i.e. admit or drop, \newline
  $\mathsf{P}(S_{t+1}|S_t,A_t)$: transition probability to state $S_{t+1}$ from $S_t$ given $A_t$ action was taken, \newline
  $R(S_t,A_t)$: reward when action $A_t$ is taken in state $S_t$, \newline
  $\tau(S_t,A_t)$: expected transition time when action $A_t$ is taken in state $S_t$, \newline
  $X_t$: time of $t$-th packet arrival.
  \vspace{2pt}
}
\captionof{table}{Notation to define the AQM Problem in Sect.~\ref{sec:model}}
\label{tab:notation}
\end{minipage}
}}
\vspace{4pt}

\section{AQM as an Optimal Decision Problem}
\label{sec:model}

In the following, we formulate the AQM problem as finding an optimal policy of a Semi Markov Decision Process where the underlying  system is essentially an AQM-capable router or switch that carries IP traffic.
We recognize that the problem can be framed in various ways depending upon the flow RTT and flow properties. 
Such distinctions, e.g., with respect to flow RTT can naturally arise during intra-datacenter or inter-datacenter communications as depicted in Fig. \ref{fig:dataCentre}.
Under the scenario of negligible RTT, we allow the arrival process to have a more general form whereas, keeping tractability in sight, a simpler model is adopted for the case with non-negligible RTT. 
In this section, the former is described in greater detail and we take up the latter in Sect.~\ref{sec:rttModel}.

In our framework where negligible RTT is assumed, the buffer is observed at arrival instants\footnote{We use instant and epoch interchangeably} and we aim to find out the optimal policy, i.e., whether it is ideal to drop or admit a packet given the state of the system. 
Apart from the action, the state evolution of the system is influenced by the packet interarrival time and the service time distributions. 
We formulate the problem using a model of the packet data arrivals that is given by gamma distributed interarrival times where the parameters of the distribution depend on the history. 
Our choice of distribution for the interarrival times allows us to fit a wide class of observed traffic flows to the model. 
Further, the service times are assumed to be exponentially distributed.
Before formalizing the system description and the framework in general, we emphasize that optimality here is defined in terms of expected long term average reward.
The expected long term average reward can be thought of as the long-term accumulation rate of instant rewards where the instant reward can be specified precisely depending upon the intended objective such as a function of throughput and/or delay.
Specifically, we formulate the reward function to capture the immediate gain in relative throughput while the given delay threshold is adhered to. 
The required notations to formally define the AQM problem are introduced in \Cref{tab:notation}.

Further, we denote the state space by $\mathcal{S}$ and the action space by $\mathcal{A} = \{0,1\}$ where $0$ and $1$ denote  admitting and dropping of a packet, respectively. These notations are used across sections with minor variation which is mentioned in respective contexts\footnote{In Sect.~\ref{sec:rttModel} (non-negligible RTT case), $\beta_t$ denotes the rate parameter of the exponentially distributed interarrival time \emph{at} $t$-th arrival.}. 

Looking at the consequence of available actions, we see that admitting a packet causes the queue length to increase by one and the effective arrival rate of the TCP flow that is given as $\beta_t/\alpha$ increases to $f_u(\beta_t/\alpha)$ immediately\footnote{In contrast, the change in arrival rate occurs with a lag when RTT is non-negligible; see Sect.~\ref{sec:rttModel} for details.}. Here, the function $f_u$ is dependent upon the exact TCP congestion control algorithm. For example, for an additive increase multiplicative decrease (AIMD) algorithm~\cite{chiu1989analysis}, $f_u(\beta_t/\alpha) = \beta_t/\alpha + a$, for some $a>0$. 
For tractability, we assume the shape parameter $\alpha$ remains fixed and vary the rate parameter $\beta$ appropriately to reflect this change.
This implies that following an admission action, $\beta_{t+1}$ assumes the value $\alpha f_u(\beta_t/\alpha)$.

In contrast, packet drops evidently do not change the queue length although the effective arrival rate $\beta_t/\alpha$ drops to $f_d(\beta_t/\alpha)$ immediately where $f_d$ is dependent upon the exact TCP congestion control algorithm. 
Again, for an AIMD algorithm, $f_d(\beta_t/\alpha) = b \beta_t/\alpha$, for some $0<b<1$. As before, we assume the shape parameter remains fixed and take $\beta_{t+1} = b \beta_{t}$. For the sake of simplicity, we derive our results with the simplest version of AIMD algorithm which uses $a = 1$ and $b = 1/2$. Our results remain valid under a wide class of elementary congestion control functions and can be obtained by replacing the increment and decrement of flow arrival rate with corresponding $f_u$ and $f_d$.

To formalize the state evolution under the given AIMD algorithm, the action $A_t=0$ causes the queue length to increase to $(Q_t+1)$ instantaneously and the possible states in the next arrival epoch could be any of the elements of the set: $\mathcal{Q}_0 = \{(q,\beta_t+\alpha):0 \le q \le Q_t+1\}$. That is, $S_{t+1} \in \mathcal{Q}_0$ and $\mathsf{P}((q,\beta_t+\alpha)|S_t,0)$ is the probability that exactly $Q_t+1-q$ many packets are served until next packet arrival.
Similarly, for the action $A_t=1$, we have $\mathcal{Q}_1 = \{(q,\beta_t/2):0 \le q \le Q_t\}$ and $\mathsf{P}((q,\beta_t/2)|S_t,1)$ is the probability of serving $Q_t-q$ many packets until the next packet arrival. 
The transition probabilities to any state outside these designated sets are zero.
Next, we derive the expression for state transition probabilities which dictate the pattern of transition between the system states. These transition probabilities are crucial inputs to the optimal policy derivation process. \looseness = -1
\vspace{-3pt}

\subsection{State Transition Probabilities}

Recall that the system state is described as the vector consisting of the current queue length and the arrival rate, i.e., $S_t = (Q_t, \beta_t)$. We are interested in deriving an expression for the transition probabilities $\mathsf{P}(S_{t+1}|S_t, A_t)$ as these are inputs to the Bellman operator which iteratively determines the value of each state. The value of each system state in turn determines the policy; see Chap. $7$ of~\cite{tijms2003first} for details. To that end, we state the following lemma.

\begin{lemma}\label{lemma:gammaRec}
For two independent Gamma random variables, $Y_{u,v} \sim \text{Gamma}(u,v)$ and $X_{w,z} \sim \text{Gamma}(w,z)$, \hfill \break $\mathsf{P}(Y_{u,v}>X_{w,z})= \mathsf{P}(Y_{u-1,v}>X_{w,z}) + \frac{\Gamma (u+w-1)}{\Gamma (u)\Gamma (w)} \big(\frac{v}{v+z}\big) ^ {u-1} \big(\frac{z}{v+z}\big) ^ w$, where $u>1$. This implies
$$\mathsf{P}(Y_{u,v}>X_{w,z})= \sum_{k=0}^{u-1} \frac{\Gamma (k+w)}{\Gamma (k+1)\Gamma (w)} \bigg(\frac{v}{v+z}\bigg) ^ k \bigg(\frac{z}{v+z}\bigg) ^ w~.$$
\end{lemma}
\begin{proof}
The proof is given in Sect.~\ref{sec:appendix}.
\end{proof}
Now, we can derive the transition probabilities as follows.

\begin{propo}\label{coro:transProbSmdp}
Given the state $S_t = (Q_t, \beta_t)$ at $t$-th arrival epoch, $Q_t$ and $\beta_t$ being the queue length and arrival rate, respectively, the transition probabilities to the state $S_{t+1} = (Q_{t+1}, \beta_{t+1})$ in the next epoch under the action $A_t=0$ are given by
 \begin{align*}
 & \quad \mathsf{P}((Q_{t+1},\beta_{t+1})|S_t,0)  = \mathbbm{1}_{\beta_{t+1} = \beta_t+\alpha}\\
 & \frac{\Gamma (Q_t+1-Q_{t+1}+\alpha)}{\Gamma (Q_t+2-Q_{t+1})\Gamma (\alpha)} \!\bigg(\frac{\mu}{\mu+\beta_{t+1}}\bigg)^{Q_t+1-Q_{t+1}}\! \!\bigg(\frac{\beta_{t+1}}{\mu+\beta_{t+1}}\bigg)^{\alpha}, 
  \end{align*}
 for $1 \le Q_{t+1} \le Q_t+1$ and 
 \begin{align*}
 & \quad \mathsf{P}((0,\beta_{t+1})|S_t,0) = \mathbbm{1}_{\beta_{t+1} = \beta_t/2}  \\
 & \bigg(1\! -\sum_{k=0}^{Q_t} \frac{\Gamma (k+\alpha)}{\Gamma (k+1)\Gamma (\alpha)} \!\bigg(\frac{\mu}{\mu+\beta_{t+1}}\bigg) ^ {Q_t+1-k} \!\bigg(\frac{\beta_{t+1}}{\mu+\beta_{t+1}}\bigg) ^ {\alpha}\bigg).
  \end{align*}
\end{propo}
\begin{proof}
See Sect.~\ref{sec:appendix}.
\end{proof}

The transition probabilities for the case when a packet is dropped are derived similarly. Note that only the rate parameter of the interarrival time variable changes to $\beta_t/2$ due to the drop and the possible queue length in the next decision epoch is at most $Q_t$. Thus, 
\begin{align*}
& \mathsf{P}((Q_{t+1},\beta_{t+1})|S_t,1) = \mathbbm{1}_{\beta_{t+1} = \beta_t/2}\\
& \thinspace \frac{\Gamma (Q_t-Q_{t+1}+\alpha)}{\Gamma (Q_t+1-Q_{t+1})\Gamma (\alpha)} \!\bigg(\frac{\mu}{\mu+\beta_{t+1}}\bigg) ^ {Q_t-Q_{t+1}} \!\bigg(\frac{\beta_{t+1}}{\mu+\beta_{t+1}}\bigg) ^ {\alpha}~, 
\end{align*}
 for $1 \le Q_{t+1}  \le Q_t$ and 
\begin{align*}
& \mathsf{P}((0,\beta_{t+1})|S_t,1) = \mathbbm{1}_{\beta_{t+1} = \beta_t/2}\\
&\thinspace \bigg(1 - \sum_{k=0}^{Q_t-1} \frac{\Gamma (k+\alpha)}{\Gamma (k+1)\Gamma (\alpha)} \bigg(\frac{\mu}{\mu+\beta_{t+1}}\bigg) ^ {Q_t-k} \bigg(\frac{\beta_{t+1}}{\mu+\beta_{t+1}}\bigg) ^ {\alpha}\bigg)~. 
\end{align*}

\begin{figure}[t!]
	\centering
	\includegraphics[width=1\linewidth]{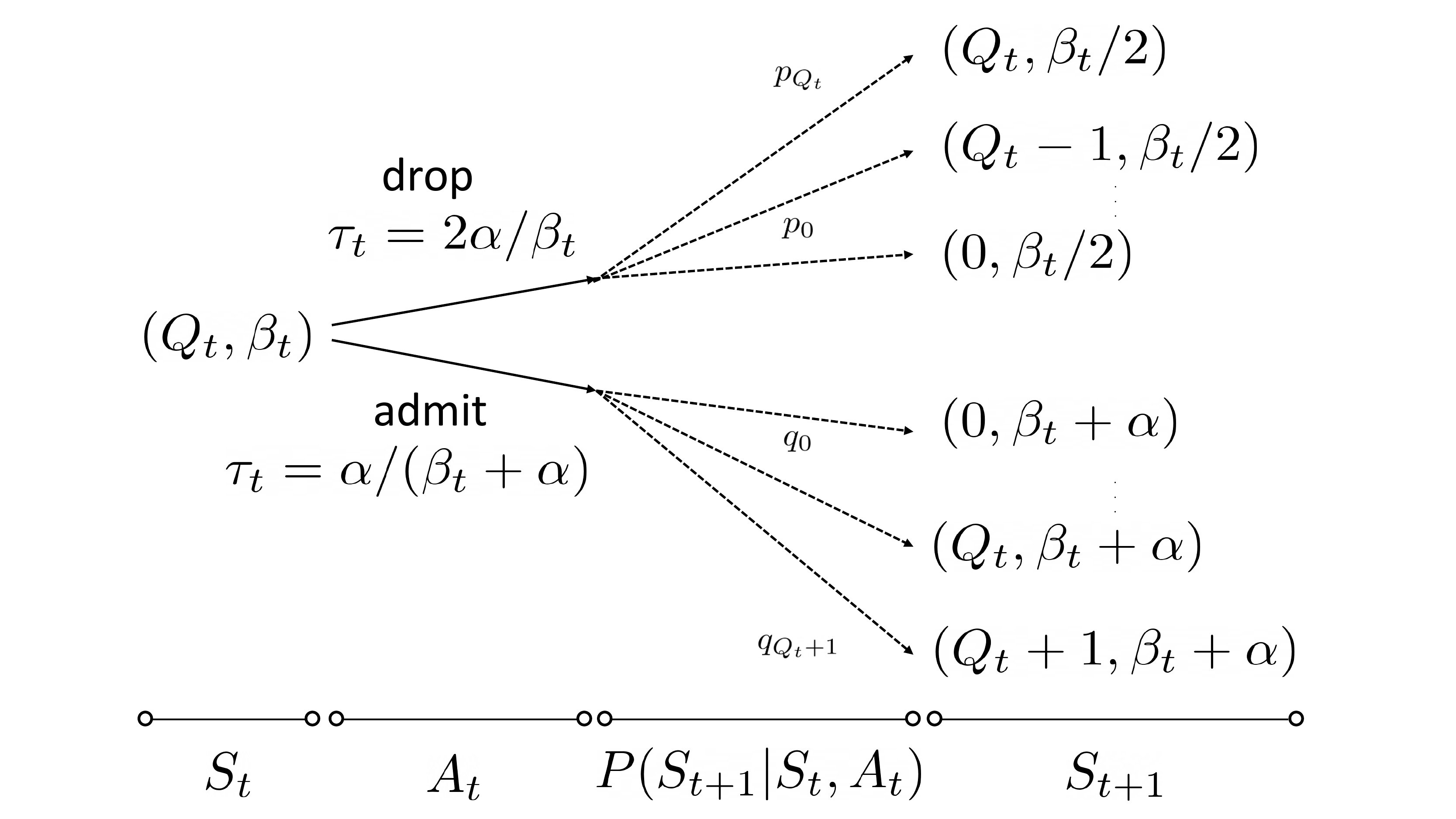}
    \caption{Starting from a system state $(Q_t, \beta_t)$, i.e., the current buffer filling and the current arrival rate parameter, the figure shows the state transition after packet admit/drop action. Solid arrows represent actions, whereas dotted arrows signify subsequent state transitions annotated with corresponding probabilities. The expected time until the next decision, i.e., the mean interarrival time is denoted by $\tau_t$.
    }
    \label{fig:stateTransitionSmdp}
    \vspace{-5pt}
\end{figure}

Unlike the MDP framework, the expected time between two decision epochs plays a significant  role in the SMDP framework that forms the basis of our analysis. 
Since drop/admit action affects the packet sending rate of TCP, the expected time between two arrival events varies. We see that the expected transition times are: $\tau(S_t,0) = \alpha/(\beta_t+\alpha)$ and $\tau(S_t,1) = 2 \alpha/\beta_t$, following the AIMD principle of TCP. 
We occasionally abbreviate $\tau(S_t,A_t)$ as $\tau_t$. A brief sketch of state transitions in this framework is provided in Fig.~\ref{fig:stateTransitionSmdp}. 
\subsection{Reward function}
Equipped with the transition probabilities, we now focus on choosing the reward function which shapes the objective of the policy learning process. The reward function represents the immediate incentive and the policy learning process aims to maximize the average accumulated reward in the long run. To align the learning process with our objective of maximizing throughput given a user-defined delay constraint, we use the reward function described in the following.
The reward function penalizes heavily whenever the delay constraint is breached and provides immediate incentive that equals the change in the square rate of throughput. The rationale here is to reflect the relative change rather than its absolute value.
Formally,
\vspace{-5pt}
\begin{align}\label{eq:rewardFnTput}
\begin{aligned}
    R(S_t,0) &= -M \mathbbm{1}_{\{(Q_t+1)/\mu>\eta\}} + \bigg(\sqrt{\frac{\beta_t+\alpha}{\alpha}} - \sqrt{\frac{\beta_t}{\alpha}}\bigg)~, \\
    R(S_t,1) &= -M \mathbbm{1}_{\{Q_t/\mu>\eta\}} + \bigg( \sqrt{\frac{\beta_t}{2 \alpha}} - \sqrt{\frac{\beta_t}{\alpha}}\bigg)~,
\end{aligned}
\vspace{-5pt}
\end{align}
where $\eta$ denotes a target delay threshold and $M$ is a large penalty value for breaching it.
Thus, the reward function adopts the goal of state-of-the-art heuristic AQM algorithms that take a target delay as a single input variable ~\cite{nicholsand}.

Equipped with the framework above and the reward function, we focus on finding the AQM policy using the value-iteration algorithm for SMDP's. The AQM policy runs on the buffer and takes the system state $(Q_t, \beta_t)$, essentially, the current buffer filling and the current arrival rate parameter as input and provides the action, i.e., whether to drop or admit to maximize accumulated predefined reward.

Formally, a policy $\pi$ is defined as a mapping $\pi:\mathcal{S} \mapsto \mathcal{A}$, i.e., given the system state in terms of buffer filling and arrival rate, the policy returns one of the possible actions, i.e., admission or dropping. Let the accumulated reward up to time $x$ be denoted by $Z(x)$, i.e., 
$Z(x) = \sum_{t =1}^{V(x)} R(S_t, A_t)$, where
$V(x) = \max \left\{t: X_t \le x\right\}$,
and $X_t$ is the arrival time of the $t$-th packet.
The expected average long-term reward is then given by:
\begin{align}
    g_i(\pi) = \lim_{x \rightarrow \infty} \frac{1}{x} \mathsf{E}_{i,\pi}[Z(x)]~,
\end{align}
where $i$ denotes the fact that the initial state was $S_i$ and $\pi$ is the used policy.

As mentioned, our objective is to find a policy $\pi$ which maximizes $g$ for each state $S_i$. To that end, we convert the SMDP problem to a discrete-time MDP using the data transformation method~\cite{tijms2003first} by appropriately scaling the rewards and transforming the transition probabilities. This is due to the fact that an SMDP with average reward criterion can be converted for the purpose of solving it to a discrete time MDP where the rewards are modified to reward accumulation rate and the transition probabilities are adjusted to reflect the changes from continuous time to discrete time transformation. Subsequently, we use the value iteration method for discrete time MDP's to determine the optimal policy. The transformed reward function $\bar{R}$ and transition probability $\bar{\mathsf{P}}$ are respectively given by: 
\begin{align} \label{eq:dataTransform}
    \begin{aligned}
    \bar{R}(S_t,A_t) &= \frac{R(S_t,A_t)}{\tau(S_t,A_t)}~,\\
    \bar{\mathsf{P}}(S_{t+1}|S_t,A_t) &= \mathsf{P}(S_{t+1}|S_t,A_t) \frac{\tau}{\tau(S_t,A_t)}, \; S_{t+1} \neq S_{t}~, \\
    \bar{\mathsf{P}}(S_{t}|S_t,A_t) &= \mathsf{P}(S_{t}|S_t,A_t) \frac{\tau}{\tau(S_t,A_t)} + 1- \frac{\tau}{\tau(S_t,A_t)}~,
    \end{aligned}
\end{align}
where $\tau$ is chosen such that $0 < \tau \le \min_{s,a} \tau(s,a)$.
\section{AQM for unknown Traffic Flow Properties}
\label{sec:inference}

In this section, we consider the problem of finding the optimal AQM policy for unknown traffic flow properties. As established in the Sect.~\ref{sec:model}, the optimal AQM policy requires accurate knowledge of the current traffic arrival rate which is determined by the initial arrival rate and the series of drop or admit actions.
Taking the current arrival rate and the queue filling as inputs, the AQM policy then prescribes whether it is optimal to drop or admit an incoming packet. 
Therefore, to propose a policy under unknown arrival characteristics, we first need to infer the respective parameters of the arrival flows.
However, estimating the model parameters while learning the optimal policy introduces the classical problem of dual control~\cite{feldbaum1960dual}. However, this can be alleviated by an exploration-exploitation Bandit-heuristic~\cite{Sutton1998}, which is computationally tractable. \looseness = -1

\subsection{Estimation of Traffic Flow Parameters}

Next, we infer the parameters governing the arrival process at data packet arrivals. To that end, we derive Maximum Likelihood Estimates (MLE) of arrival shape $\alpha$ and arrival rates $\beta_t$ as used by the parametric models in Sect.~\ref{sec:model}. 
Let $\{X_n\}$ and $\{A_n\}$ denote the sequence of arrival times and the actions respectively. Further, the corresponding packet interarrival times are denoted by $W_t$, i.e., $W_t = X_{t+1}-X_t$. Recall that $A_t=1$ signifies a packet drop and $A_t=0$ denotes packet admission. We take a tractable model of Gamma interarrival times, i.e. 
\begin{align}\label{eq:aimdGamma}
\begin{aligned}
    W_t|\alpha, \beta_t &\sim \text{Gamma}(\alpha, \beta_t)~, \\
    \beta_{t+1} &= (\beta_t+\alpha)^{1-A_t}(\frac{\beta_t}{2})^{A_t}~, 
\end{aligned}
\end{align}
following the AIMD congestion control principle of TCP. Given the sequence of interarrival times  $\mathbf{W} = \{W_n\}_{n=1}^k$ and actions $\mathbf{A} = \{A_n\}_{n=1}^k$, we aim to estimate the common shape parameter $\alpha$ and the initial rate parameter $\beta_1$ 
to derive the transition probabilities $\mathsf{P}(S_{t+1}|S_t, A_t)$ described in Sect.~\ref{sec:model} to find the optimal policy. 
The likelihood of the parameters for the observed sequence $(\mathbf{W},\mathbf{A})$ is given by
\begin{align*}
    L(\alpha, \beta_1|\mathbf{W},\mathbf{A}) \!=\! \frac{(\prod_{n=1}^k \beta_n)^\alpha (\prod_{n=1}^k W_n))^{\alpha-1} e^{-\sum_{n=1}^k \beta_n W_n}}{(\Gamma \alpha)^k}.
\end{align*}
Note that the RHS is simply product of Gamma densities and the dependence on $\mathbf{A}$ is implicit. Now, taking logarithm we obtain the log-likelihood $l(\alpha,\beta_1)$ as
\begin{align}\label{eq:likelihood}
    \begin{aligned}
    l(\alpha,\beta_1) 
    & = \alpha \sum \log \beta_n +(\alpha-1)\sum \log W_n  \\ 
    & \quad -\sum
    \beta_n W_n - k \log(\Gamma \alpha)~.
    \end{aligned}
\end{align}
To optimize the likelihood we obtain 
\begin{align}\label{eq:mlePdes}
\begin{aligned}
    \frac{\partial l}{\partial \alpha} &= \sum \log \beta_n + \sum \bigg(\frac{\alpha}{\beta_n} - W_n\bigg) \frac{\partial \beta_n}{\partial \alpha}  \\
    & \quad +\sum \log W_n - k \frac{\partial}{\partial \alpha} \log(\Gamma \alpha)~,  \\
    \frac{\partial l}{\partial \beta_1} &= \sum \bigg(\frac{\alpha}{\beta_n} - W_n\bigg) \frac{\partial \beta_n}{\partial \beta_1}~,
\end{aligned}
\end{align}
where 
\begin{align*}
    \frac{\partial \beta_n}{\partial \alpha} &= \bigg( 1 + \frac{\partial \beta_{n-1}}{\partial \alpha} \bigg)^{1-A_{n-1}} \bigg( \frac{1}{2}\frac{\partial \beta_{n-1}}{\partial \alpha}\bigg)^{A_{n-1}}~, \\
    & \qquad n = 3,4,\dots,k, \\
    \text{with} \quad \frac{\partial \beta_2}{\partial \alpha} &= \mathbbm{1}_{\{A_1=0\}} \quad \text{and} \; 0^0 = 1~, \; \text{by convention}.
\end{align*}
Further,
\begin{align*}
    \frac{\partial \beta_n}{\partial \alpha} &= \bigg(\frac{\partial \beta_{n-1}}{\partial \beta_1} \bigg)^{1-A_{n-1}} \bigg( \frac{1}{2} \frac{\partial \beta_{n-1}}{\partial \beta_1}\bigg)^{A_{n-1}} \\
    &= \bigg( \frac{1}{2} \bigg)^{A_{n-1}} \frac{\partial \beta_{n-1}}{\partial \beta_1}~, \quad n = 2,4,\dots,k~. 
\end{align*}
To get the maximum likelihood estimates 
$\hat{\alpha}$ and $\hat{\beta}_1$, we numerically find the zero point of \eqref{eq:mlePdes} that maximizes \eqref{eq:likelihood}.

\subsection{Inference under unknown TCP Congestion Control}

In case the exact congestion control algorithm of TCP is not known beforehand, we assume the effective arrival rate $\beta_{t+1}/\alpha$ in the next epoch is changed according to a polynomial function (e.g., TCP CUBIC~\cite{ha2008cubic}) of the present effective arrival rate $\beta_t/\alpha$. We can then rewrite \eqref{eq:aimdGamma} as 
\begin{align}\label{eq:aimdGammaPolynomial}
\begin{aligned}
    W_t|\alpha, \beta_t &\sim \text{Gamma}(\alpha, \beta_t)~,\\
    \frac{\beta_{t+1}}{\alpha} &= \bigg(f_u\bigg(\frac{\beta_t}{\alpha}\bigg)\bigg)^{1-A_t}\bigg(f_d\bigg(\frac{\beta_t}{\alpha}\bigg)\bigg)^{A_t}~, 
\end{aligned}
\end{align}
where $f_u$ and $f_d$ are polynomials governing the change of the arrival rate following an admit and drop, respectively. The log-likelihood has a similar form as shown above, although estimates of the coefficients of the polynomials, together with $\alpha$ and $\beta_1$, are required. This can either be done numerically or by calculating partial derivatives explicitly, similar to \eqref{eq:mlePdes}. Subsequently, one can find their zeros corresponding to the maxima, albeit with additional equations for $\partial l/\partial c_j$, for each coefficient $c_j$. For example, if $f_u$ is a cubic polynomial and $f_d$ is a linear function, this method will lead to calculation of six additional partial derivatives.

Using the parameter estimates along with the observed series of drop or admit actions, we infer the current arrival rate $\hat{\beta_t}$ which together with the current queue filling $Q_t$ describes the system state. We thus use the AQM policy similar to Sect.~\ref{sec:model} to find the optimal action.
\section{AQM under Non-negligible RTT} \label{sec:rttModel}
In this section, we focus on the case where the RTT between the sender and the receiver of a flow is non-negligible. 
As mentioned in Sect.~\ref{sec:model}, we trade model flexibility for tractability for this case. This tractability is essential given the case where the switch acts on multiple concurrent flows. We derive our policy for the single flow case in Sect.~\ref{sec:rttModelSF} whereas the multiple flow case is described in Sect.~\ref{sec:rttModelMF}. As before, our objective is to maximize the total throughput subject to a user-defined delay constraint.
\vspace{-5pt}
\subsection{Single Flow}\label{sec:rttModelSF}
In the following, we assume that both packet interarrival times and the packet service times are exponentially distributed. Further, the RTT is denoted as $r$ and we have $r>0$. We assume that an estimate of the exact flow RTT is known to the switch.
Note that the sender adapts the sending rate, i.e., the arrival rate to the switch, only after it receives a congestion signal from the receiver, which has a propagation time of $r/2$.

Thus, the delay between an action at the switch and the arrival of the congestion signal at the sender corresponds to the propagation time from the switch to the sender via the receiver, which approximately equals $r$. 
Thus an action takes effect at the switch only after a time period $r$, and all arrivals meanwhile are admitted given there is enough room in the output buffer. 
For tractability, we further assume that the arrivals and the corresponding admit actions in this period have no direct consequence on the way arrival rate is changed. Although an approximation, we expect this assumption to have no major consequence and take advantage of the corresponding simpler state representation.

\begin{figure}[t!]
	\centering
	\includegraphics[width=1\linewidth]{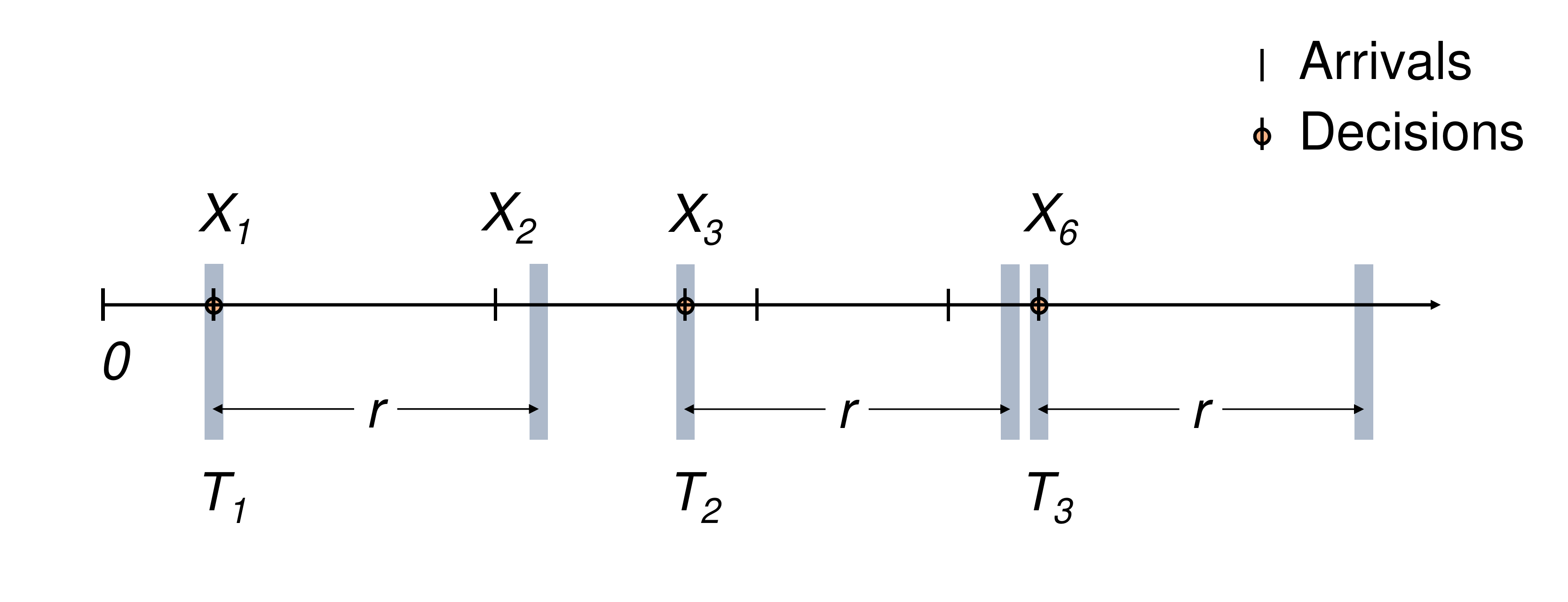}
    \caption{Arrival times $X_i$ and decision times $T_t$ are shown on the time axis. The first job arrives at time $X_1$ with rate $\beta_1$. After $X_1$, $X_2$ is the only arrival in $(X_1, X_1+r]$ and is assumed to have no bearing on the arrival rate. We see $X_3=T_2$ and the arrival rate is changed to $\beta_2$ at $T_2$ according to the action taken at $X_1$. The rate again changes to $\beta_3$ at $T_3=X_6$.
    }
    \label{fig:rttDiagram}
\end{figure}

Recall that we denote the arrival times as $\{X_s\}_{s \in \setOfNaturals}$. 
We take a decision whether to admit or drop on the first packet arrival at time $X_1$. Thereafter all arrivals in the interval $(X_1, X_1+r]$ are admitted subject to enough room in the buffer and the first arrival after $(X_1+r)$ can be potentially dropped. The arrival rate changes at this new arrival after $(X_1+r)$ according to the action taken at time $X_1$. Subsequently, the decision making process goes dormant for another time period of length $r$  and the drop/admit decision is taken again at the first arrival after this period. The change in arrival rate also takes place at this time. 
An illustrative example is given in Fig. \ref{fig:rttDiagram}. We call the arrival epochs where a drop decision can be possibly made as \textit{decision times} and denote them as $\{T_t\}_{t \in \setOfNaturals}$. As explained, $T_i$'s ($i \ge 2$) are also the time points where the arrival rate changes according to the action taken at $T_{i-1}$.

The service times are assumed independently and identically distributed (iid) as Exp$(\mu)$ and given the relevant arrival rate parameter $\beta_k$, we have $$X_{i+1}-X_i \sim \text{Exp}(\beta_k),~ i \in \setOfNaturals,$$
for some $k \in \setOfNaturals$. Note that, unlike Sect.~\ref{sec:model}, the arrival rates do not change at every arrival. Rather \textit{the change happens at the decision times} owing to the delay factor of $r$ and the fact that intermediate arrivals do not influence the arrival rate directly. Focusing on the inter-decision time, we observe that the residual time to the next decision time after one RTT $r$ is also distributed as Exp$(\beta_j)$, for some $j \in \setOfNaturals$, following the memoryless property of exponential distribution. Further, the change in the arrival rate $\beta$ happens only at the decision times. Therefore, we know for the inter-decision times
\begin{align}\label{eq:rttTimeDist}
T_{i+1}-T_i|\beta_i \sim r+\text{Exp}(\beta_i),~ i \in \setOfNaturals~,
\end{align}
as $\beta_i$ is the arrival rate during the interval $[T_i, T_{i+1})$. 

To derive the optimal decision at times points $T_i$, we adopt the SMDP framework similar to the derivations of Sect.~\ref{sec:model}. We retain much of our notations from Sect.~\ref{sec:model} with the exception of $\beta_t$ which denotes the rate parameter of packet interarrival times \emph{at} $t$-th decision epoch. In contrast to Sect.~\ref{sec:model}, $\beta_t$ in this case is already known at $(t-1)$-th arrival epoch and, hence, is a valid input to the policy.

Note that in the time interval $(T_i, T_i+r]$, the queue length grows as the population of a birth-death process, where the transition matrix $\mathsf{P}(s)$ over a time interval of length $s$ is given by $e^{s G_j}$. 
Here, $G_j$ is the $(L+1) \times (L+1)$ intensity matrix of the process in the relevant interval where $L$ is the buffer size. It is given by 
{\small
\begin{equation*}
G_j = 
\begin{bmatrix}
-\beta_j & \beta_j & 0 & \dots & \dots\\
\mu & -\mu-\beta_j & \beta_j & 0 & \dots\\
0 & \mu & -\mu-\beta_j & \beta_j & \dots\\
\dots & \dots & \dots & \dots & \dots\\
0 & \dots & \dots & \mu & -\mu
\end{bmatrix},    
\end{equation*}
}
\noindent for some $j$. Further, in the interval $(T_i+r, T_{i+1})$ there can only be packet departures and hence the transition matrix for any interval of length $s$ is given by $e^{s G}$ where
{\small
\begin{align} \label{eq:G0}
    G = 
\begin{bmatrix}
0 & 0 & 0 & \dots & \dots\\
\mu & -\mu & 0 & 0 & \dots\\
0 & \mu & -\mu & 0 & \dots\\
\dots & \dots & \dots & \dots & \dots\\
0 & \dots & \dots & \mu & -\mu
\end{bmatrix}.    
\end{align}
}
The following fact helps us derive the transition matrix between two consecutive decision epochs.
\begin{fact}\label{lemma:expmIntegral}
If all eigenvalues of the matrix $A$ are positive, 
$$ \int_0^{\infty} e^{-uA}du = A^{-1}~.$$
\end{fact}
\begin{proof}
This well-known result follows from the fact that $\lim_{t \to \infty} e^{-t A} = 0$ for a positive definite matrix $A$.
\end{proof}
We now derive the transition matrix for the queue length over the interval between two consecutive decision epochs. The first step towards this is to find the transition probabilities for the queue length as the change of arrival rate following a transition is deterministic. \looseness= -1

\begin{propo}\label{thm:transitionProbRtt}
The transition probability matrix $P_i$  for queue length Q over the interval $(T_i, T_{i+1})$ is given by
$$ P_i = \beta_i e^{r G_i} (\beta_i I-G)^{-1}.$$
\end{propo}
\begin{proof}
See Sect.~\ref{sec:appendix}.
\end{proof}

Observe that $\mathsf{P}(Q_{t+1}|Q_t,1)$ corresponds to the $Q_t$-th row of the transition matrix over the interval $[T_t,T_{t+1})$. Under $A_t = 1$, the transition matrices over intervals $[T_t,T_{t+1})$ and $(T_t,T_{t+1})$ are identical as the incoming packet is dropped. Hence, $\mathsf{P}(Q_{t+1}|Q_t,1)$ is given by the $(Q_t,Q_{t+1})$-th element of $P_t$. However, under $A_t = 0$, the queue length jumps to $\max(Q_t+1,L)$ at time $T_t$ and the the $(Q_t,\max(Q_t+1,L))$-th element represents $\mathsf{P}(Q_{t+1}|Q_t,0)$. Finally, the state transition probabilities can be expressed as follows:
\begin{align} \label{eq:transitionRtt}
\begin{aligned}
    \mathsf{P}(S_{t+1}|S_t,0) &= \mathsf{P}(Q_{t+1}|Q_t,0) \mathbbm{1}_{\beta_{t+1} = \beta_t+1},  \\
    \mathsf{P}(S_{t+1}|S_t,1) &= \mathsf{P}(Q_{t+1}|Q_t,1) \mathbbm{1}_{\beta_{t+1} = \beta_t/2}.
\end{aligned}
\end{align}
Further, similar to \eqref{eq:rewardFnTput}, we define the reward function as:
\begin{align}\label{eq:rewardRtt}
\begin{aligned}
    R(S_t,0) &= -M \mathbbm{1}_{\{(Q_t+1)/\mu>\eta\}} + \bigg(\sqrt{\beta_{t}+1} - \sqrt{\beta_{t}}\bigg), \\
    R(S_t,1) &= -M \mathbbm{1}_{\{Q_t/\mu>\eta\}} + \bigg( \sqrt{\beta_{t}/2} - \sqrt{\beta_{t}}\bigg)~.
\end{aligned}
\end{align}
Also, it is immediate from \eqref{eq:rttTimeDist} that the expected inter-decision time $\tau(S_t,A_t) = r+ 1/\beta_t$. 
Equipped with the transition probabilities, reward function and expected transition times, we can now transform the SMDP problem into a discrete time MDP problem like Sect.~\ref{sec:model} and perform value iteration on the transformed MDP to find out the best policy.

\subsection{Multiple Flows} \label{sec:rttModelMF}
In this section, we extend our SMDP formulation from Sect.~\ref{sec:rttModelSF} in presence of concurrent flows with different RTT. 
We assume that the flow RTT's are known or estimated  at the switch running AQM. \looseness = -1 

We define the decision times for a flow of interest as the instances of the first packet arrivals that are at least an RTT apart.
That is for a fixed flow, when an AQM decision is carried out for one packet, the subsequent packets within one RTT window are admitted. 
The next AQM decision is taken for the first packet after the RTT window has passed.
We will see in the following that, unlike the single flow case, the time since the last decision epoch involving a packet from each flow plays an important role in formalizing the framework here. \looseness = -1 

Let us denote the packet arrival times for the $j$-th flow as $\{X_{js}\}$, with packet index $s \in \setOfNaturals$ and, similar to the single flow case, we assume for the $j$th flow
$$ X_{j (i+1)}-X_{ji}|\beta_{jk} \sim \text{Exp}(\beta_{jk}),~ i \in \setOfNaturals,$$
for some $k \in \setOfNaturals$, $j \in [n]$ and $[n] =\{1,2,\dots, n\}$, $n$ being the number of concurrent flows. 
The decision times are denoted as $\{T_{s}\}_{s \in \setOfNaturals}$ and we see that the mean arrival rate for each flow remains constant between two decision epochs. 
This is because we assume that the AQM signal reaches the sender after an RTT. 

\begin{figure}[t!]
	\centering
	\includegraphics[width=1\linewidth]{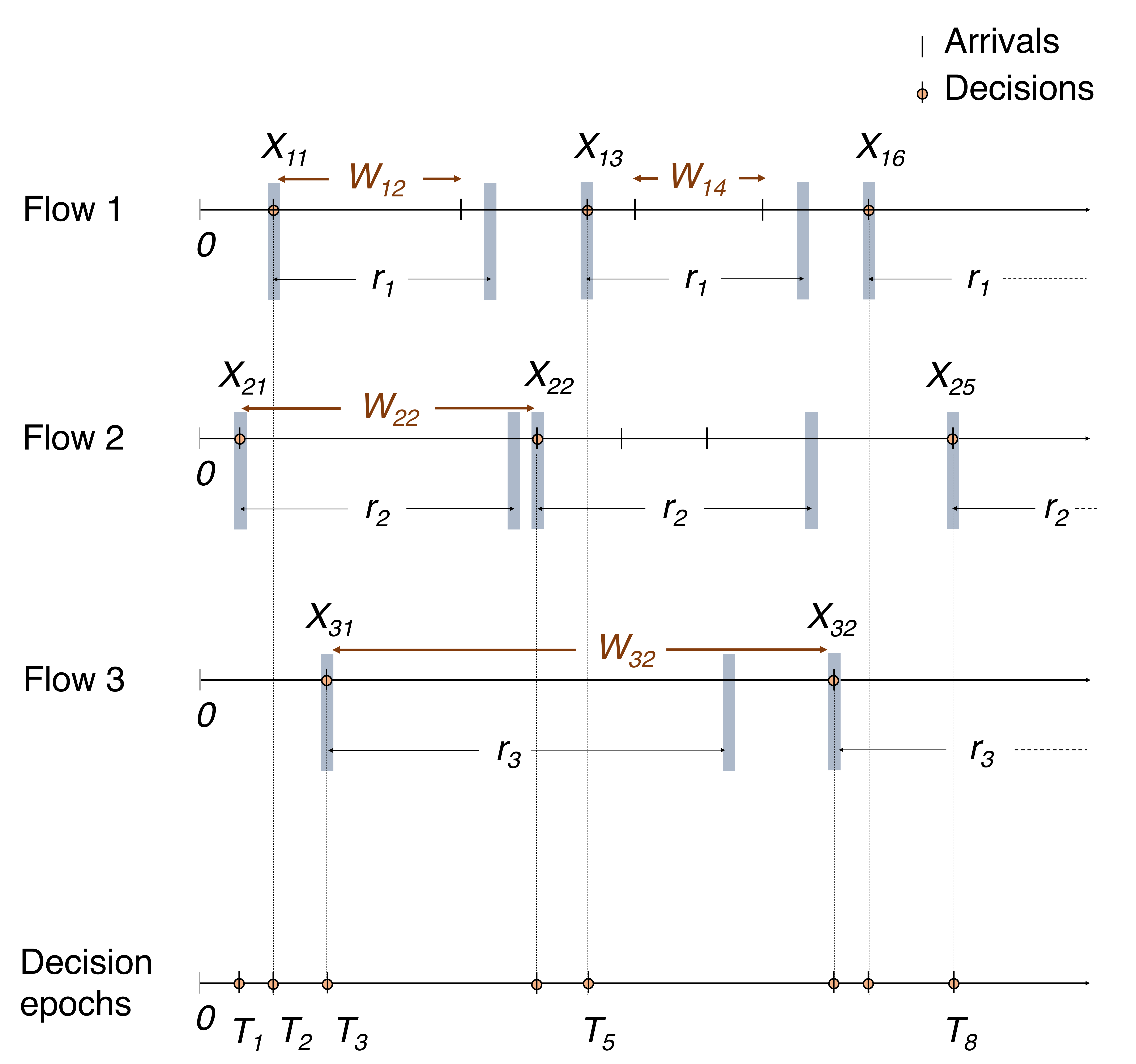}
    \caption{Illustration of decision times when three flows pass through the switch buffered link. The first decision time is the first arrival, which takes place at $T_1 = X_{21}$. Next decision time concerning flow $2$ cannot be earlier than $X_{21}+r_2$. The first arrival from flow $1$ or $3$ can potentially be the next decision time, which turns out to be $T_2 = X_{11}$. Likewise, the decision process on flow $1$ goes dormant in $(X_1, X_1+r]$. Note that decision times are the times where the arrival rates change for the associated flow. The interarrival time $W_{ij}$ is exponentially distributed with a parameter that was fixed at the last decision time concerning flow $i$, according to the action taken at the second last decision time on that flow.}
    \label{fig:rttDiagramMF}
\end{figure}

Let $\pmb{\beta}_t = (\beta_{1t}, \beta_{2t}, \dots, \beta_{nt})^T$ be the parameter vector of packet interarrival times at the $t$-th decision epoch and $\pmb{r} = (r_{1}, r_{2}, \dots, r_{n})^T$ denote the vector of flow RTTs. 
Further, we express the \emph{age} vector at $t$-th decision epoch by $\pmb{u}_t = (u_{1t}, u_{2t}, \dots, u_{nt})^T$ where $u_{jt}$ denotes the time since the last decision epoch concerning a packet from the $j$-th flow. 
In absence of a last decision time concerning flow $j$, i.e., before the arrival of the first packet from flow $j$, we define $u_{jt} = r_j$. 
Observe that if the $t$-th decision epoch involves a packet from flow $j$, we have $u_{jt} = 0$. 
An illustration of the decision epochs for the case of three concurrent flows is provided in Fig.~\ref{fig:rttDiagramMF}.

At the $t$-th decision epoch, let $Y_{jt}$ denote the time until the next AQM decision on flow $j$. In other words, $Y_{jt}$ indicates the time until the arrival of a packet from flow $j$ that is at least one RTT after the last decision epoch from flow $j$. 
Given the RTT $r_j$ of flow $j$, we can write 
$$Y_{jt} = W_j+\max(0,r_j-u_{jt})~,$$
where $W_j$ denotes the residual time to the packet arrival from $j$-th flow after $r_j$ amount of time has passed since the last decision epoch involving flow $j$. By the memoryless property of exponential distribution we know that $W_j \sim \text{Exp}(\beta_{jt})$, for $j \in [n]$. 
Thus, the time until the $(t+1)$-st decision epoch across all flows is given by
\begin{align} \label{eq:interDecisionMF}
    Y_t = \min_{j \in [n]} Y_{jt}~.    
\end{align}

Therefore, we can write the inter-decision time $T_{t+1}-T_t = Y_t$ as 
$$Y_t|\pmb{\beta}_t, \pmb{u}_t \sim \min_{j \in [n]}\{W_j+\max(0,r_j-u_{jt})\},~ t \in \setOfNaturals.$$
Equipped with the description above we now focus on formalizing the SMDP to derive the optimal AQM action corresponding to any given switch state. 
Apart from the individual flow arrival rates and the queue length, the state description in this case includes the \emph{age} vector $\pmb{u}_t$ and the flow index of the incoming packet. 
To formulate the AQM problem as an SMDP, we use the notations introduced in Sect.~\ref{sec:model} with the following exceptions:
\begin{itemize}
  \item[] $\pmb{\beta}_t$: flow-wise rate parameter \emph{vector} of packet interarrival times \emph{at} the $t$-th decision epoch,
  \item[] $\pmb{u}_t$: flow-wise age \textit{vector}, i.e., time since last decision epoch involving a packet from the concerned flow at the $t$-th decision epoch,
  \item[] $S_t$: state of the system observed by the packet corresponding to the $t$-th decision epoch, given by $(j, Q_t, \pmb{\beta}_t, \pmb{u}_t)$ where $j$ denotes the flow index of the packet for which an AQM decision is to be taken.
\end{itemize}

With the flow-wise age $\pmb{u}_t$ being a continuous variable  we derive next the corresponding transition kernels.
Let $B = (j, Q_{t+1}, \pmb{\beta}_{t+1}, U)$ denote the set that the system state at $t+1$, i.e., $S_{t+1}$, can possibly belong to. 
As usual, we restrict our focus on $U \in \mathcal{B}(\setOfReals^n)$, the Borel sigma algebra over $\setOfReals^n$.
Due to the known fact that right semi-closed rectangles generate $\mathcal{B}(\setOfReals^n)$, it is sufficient to consider the sets $U$ having the form $U = I_1 \times I_2 \times \dots I_n$ where each $I_j$ is a right semi-closed interval. \looseness = -1

\begin{figure*}[b]
\noindent\rule{\textwidth}{1pt}
\begin{align} \label{eq:transKernelFactorization}
   G_j(a,b)  &=\mathbbm{1}_{j\in \{i_1 \dots i_l\}} \prod \limits_{1 \le k \le l} e^{\big((c_{i_k}-c_{i_{k-1}}) H_{i_k} \big)}
   \int_a^b e^{\big(\sum_{1 \le k \le l, i_k \ne j}-\beta_{i_k}(y- c_{i_k})\big)} e^{(y-c_{i_l}) H_{i_{l+1}}} \beta_j e^{-\beta_{j}(y- c_{j})} dy  \nonumber\\
   & = \mathbbm{1}_{j\in \{i_1 \dots i_l\}} \beta_j e^{\left(\sum\limits_{1 \le k \le l} \beta_{i_k} c_{i_k}\right)} \prod \limits_{1 \le k \le l} e^{(c_{i_k}-c_{i_{k-1}}) H_{i_k}-c_{i_l} H_{i_{l+1}}}
   \int_a^b e^{\big(\sum_{1 \le k \le l}-\beta_{i_k} y\big)}e^{y H_{i_{l+1}}} dy \tag{$\ast$} \\
   & = \mathbbm{1}_{j\in \{i_1 \dots i_l\}} \beta_j e^{\left(\sum\limits_{1 \le k \le l} \beta_{i_k} c_{i_k}\right)}   \prod \limits_{1 \le k \le l} e^{(c_{i_k}-c_{i_{k-1}}) H_{i_k}-c_{i_l} H_{i_{l+1}}}
   \left(\sum_{1 \le k \le l}\beta_{i_k}I -H_{i_{l+1}}\right)^{-1} \nonumber \\
   & \quad \bigg(e^{a \big(\sum_{1 \le k \le l}\beta_{i_k}I -H_{i_{l+1}}\big)}- e^{b \big(\sum_{1 \le k \le l}\beta_{i_k}I -H_{i_{l+1}}\big)}\bigg)~, \nonumber 
\end{align}
\end{figure*}

Going back to \eqref{eq:interDecisionMF}, we see that the value of the age vector at the decision epoch $(t+1)$ is given by $\pmb{u}_t+Y_t$. 
This motivates us to define the translates
$$I_j' = \{x:x+u_{jt} \in I_j\}\;, \; j \in [n]~.$$
Subsequently, the transition kernel can be written as:
\begin{align}\label{eq:transitionKernelRttMF}
\begin{aligned}
    &\mathsf{P}(S_{t+1}\in B|S_t, 0) = \mathbbm{1}_{\pmb{\beta}_{t+1} = \pmb{\beta}_{t}+e_j} \\ 
    & ~ \mathsf{P} \bigg(Y_t \in \bigcap\limits_{k \in [n]}I_k', Y_t = Y_{jt}, V_t = Q_{t+1}-Q_t | A_t = 0\bigg) , \\ 
    &\mathsf{P}(S_{t+1}\in B|S_t, 1) = \mathbbm{1}_{\pmb{\beta}_{t+1} = h(\pmb{\beta}_{t},j)}  \\
    &  ~ \mathsf{P} \bigg(\!Y_t \in\!\!\! \bigcap\limits_{k \in [n]}\!\!I_k', Y_t \!=\! Y_{kt}, V_t \!=\! Q_{t+1}\!-\!Q_t | A_t \!=\! 1\!\bigg)\!~,
\end{aligned}
\end{align}
where $V_t$ denotes the change in queue length in the interval $[T_t, T_{t+1})$ and $h(\pmb{x},j) = (x_1,x_2,\dots,x_i/2,\dots,x_n)$ and $e_j$ denotes the unit vector whose $j$-th coordinate equals $1$. Essentially, the indicator function involving $\pmb{\beta}_{t}$ in the RHS of \eqref{eq:transitionKernelRttMF} denotes the fact that the arrival rate can only change to specific values which are determined by the drop or admit action and the AIMD principle of TCP. Further, $(t+1)$-th decision epoch involves flow $j$ iff the corresponding inter-decision time is $Y_t = Y_{jt}$. Finally, the age vector $\pmb{u}_{t+1}$ in the next epoch belongs to the set $U$ iff $Y_t \in I_k- u_{kt}$, $1 \le k \le n$.
We further introduce the following shorthand notations:
\begin{align*}
    \bigcap\limits_{j \in [n]}I_j' = (a,b], ~& \text{when} ~ \bigcap\limits_{j \in [n]}I_j \ne \phi ~, \\
    \max(0,r_j-u_{jt}) &:= c_j, \quad j \in [n]~.
\end{align*}
Given a decision at time zero we use $c_j$ to denote the time span during which there cannot be a decision concerning flow~$j$. In particular, this time span equals the RTT as $c_i = r_i$, since $S_t = \{i, Q_t, \pmb{\beta}_t, \pmb{u}_t\}$. 
Now lets order the time spans as $c_{i_1} \le c_{i_2} \le \dots \le c_{i_n}$. Evidently, for $(a,b] \subset (-\infty, c_{i_1}]$, we have for the time to the next decision of the flow $i_1$ that $\mathsf{P}(Y_t \in (a,b]) = 0$. 
Further, for $(a,b] \subset (c_{i_l}, c_{i_{l+1}}]$ we can express the probability of the time to the next decision falling in this time span as~\eqref{eq:transKernelFactorization},
where we have used the shorthand:
$$G_j(a,b) = P (Y_t \in (a,b], Y_t = Y_{jt}, V_t = Q_{t+1}-Q_t)~.$$
See Sect.~\ref{sec:appendix} for proof.

From~\eqref{eq:transKernelFactorization}, the probability under the drop action $A_t = 1$ is given by the $(Q_t,Q_{t+1})$-th element of the kernel whereas the $(Q_t,\max(Q_t+1,L))$-th gives the required probability under the admit action $A_t = 0$.
Finally, similar to the single flow case, we define the reward function $R(S_t,A_t)$ as 

\begin{align}\label{eq:rewardMultFlows}
\begin{aligned}
    R(S_t,0) &= -M \mathbbm{1}_{\{(Q_t+1)/\mu>\eta\}} + \bigg(\sqrt{\beta_{j}+1} - \sqrt{\beta_{j}}\bigg),  \\
    R(S_t,1) &= -M \mathbbm{1}_{\{Q_t/\mu>\eta\}} + \bigg(\sqrt{\beta_{j}/2} - \sqrt{\beta_{j}}\bigg)~.
\end{aligned}
\end{align}
To derive the optimal policy, we can now simulate the trajectory of the system according to \eqref{eq:transKernelFactorization} and use a function approximator, e.g., DQN~\cite{mnih2015human} to learn the Q-values for each state-action pair. Subsequently, the optimal policy is given by the action with higher Q-value for each state. The advantage of our approach over model-free learning is that it requires a lot less data to reliably predict the optimal action.
\section{Evaluation}
\label{sec:numerical}
\begin{figure}[t]
	\centering
	\begin{subfigure}[t]{0.48\columnwidth}
		\centering
		\includegraphics[width=1\textwidth]{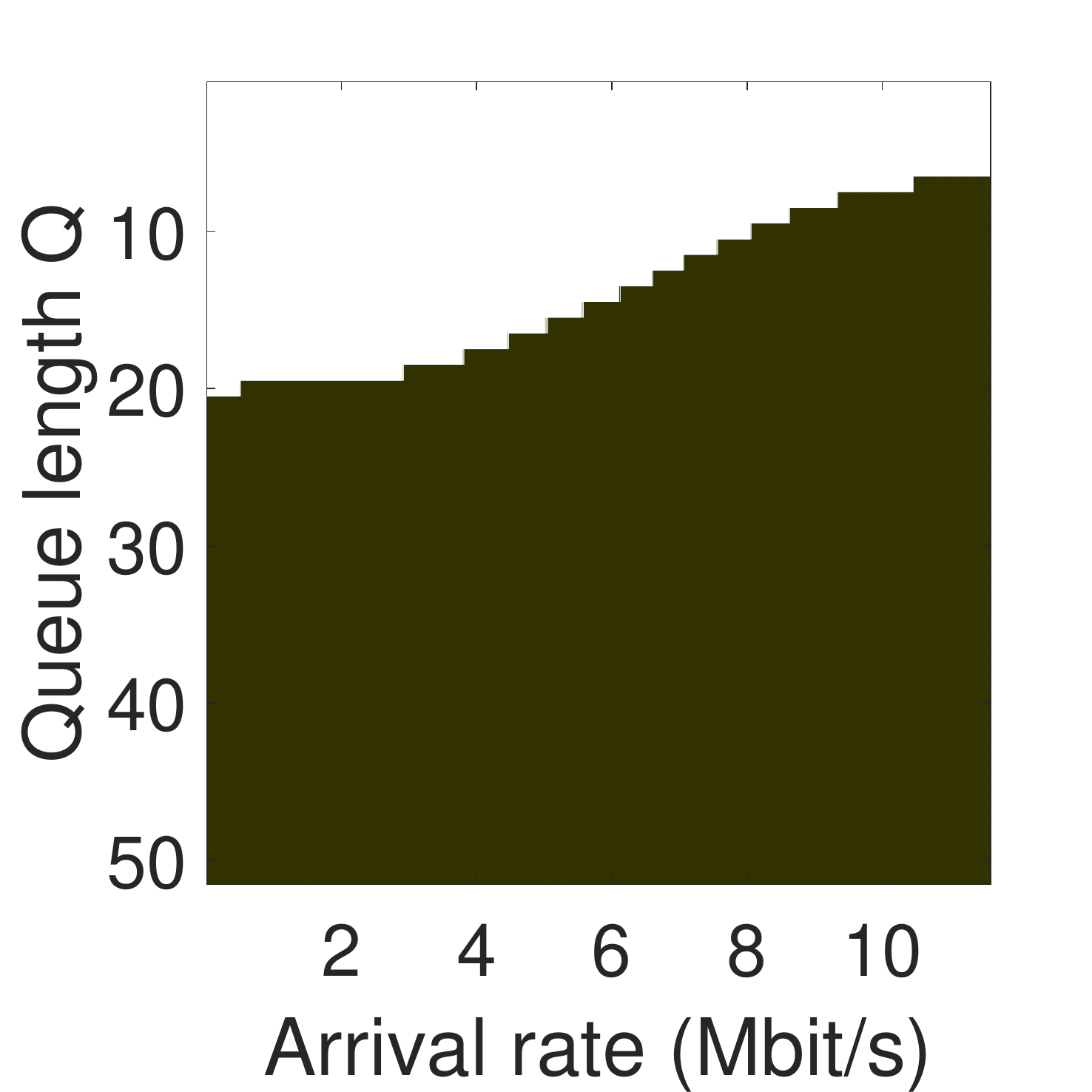}
		\caption{\label{fig:optPolicyLow}%
		Service rate = $5$ Mbit/s}
	\end{subfigure}
    \hfill
	\begin{subfigure}[t]{0.48\columnwidth}
		\centering
		\includegraphics[width=1\textwidth]{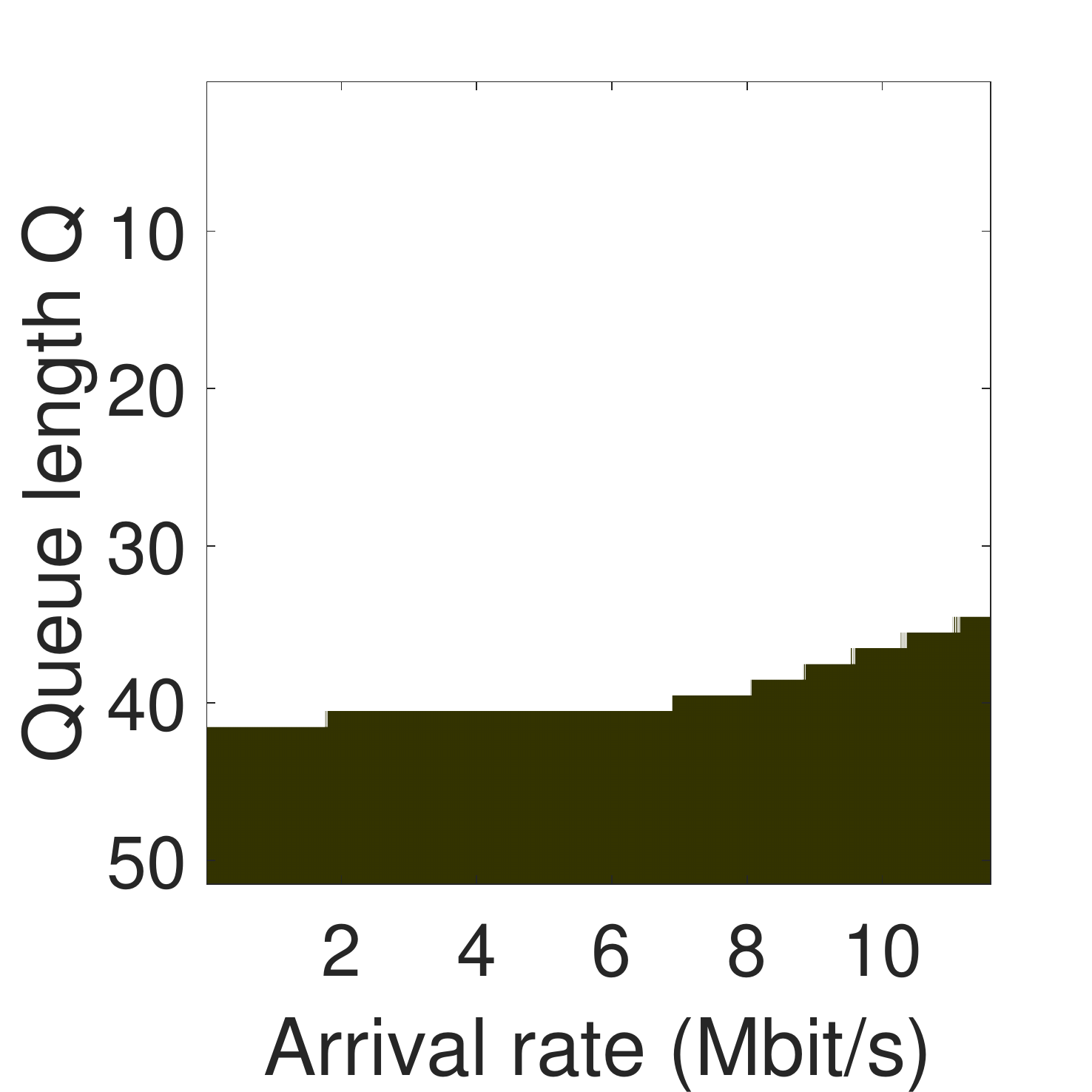}
		\caption{\label{fig:optPolicyHigh}%
        Service rate = $10$ Mbit/s}
	\end{subfigure}
\caption{\label{fig:optPolicy}%
PAQMAN under negligible RTT for two different service rates: the plot considers arrival rates~$\in [0.01,12]$~Mbit/s and the buffer size is set to $50$ packets. The target delay threshold $\eta = 50$~ms corresponds to $20$ packets in (a) and to $40$ packets in (b). The darker region shows where an incoming packet should be dropped.}%
\end{figure}

\begin{figure}[t]
	\centering
	\begin{subfigure}[t]{0.48\columnwidth}
		\centering
		\includegraphics[width=1\textwidth]{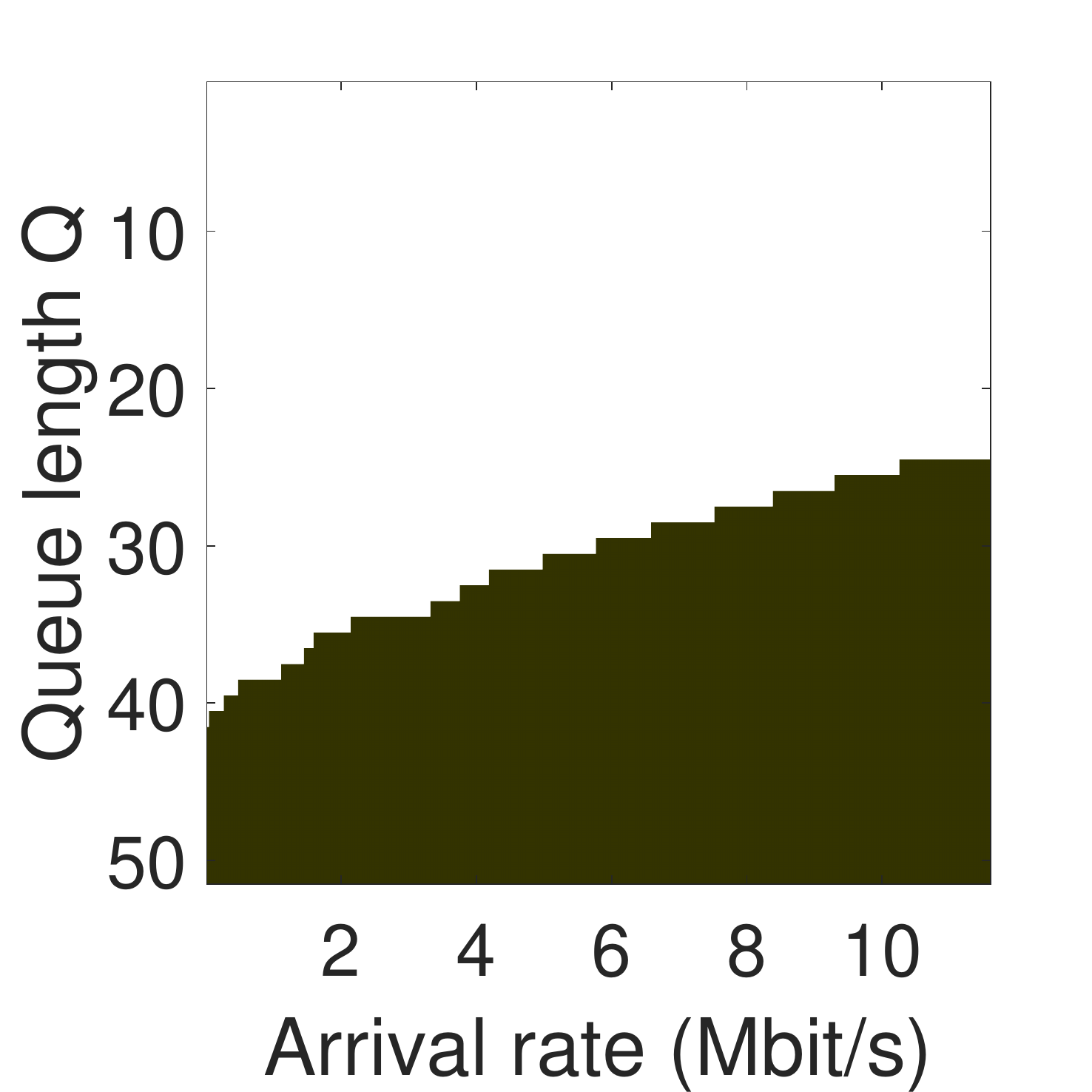}
		\caption{\label{fig:optPolicyLowRtt2}%
		RTT = $2$ ms}
	\end{subfigure}
    \hspace{2pt}
	\begin{subfigure}[t]{0.48\columnwidth}
		\centering
		\includegraphics[width=1\textwidth]{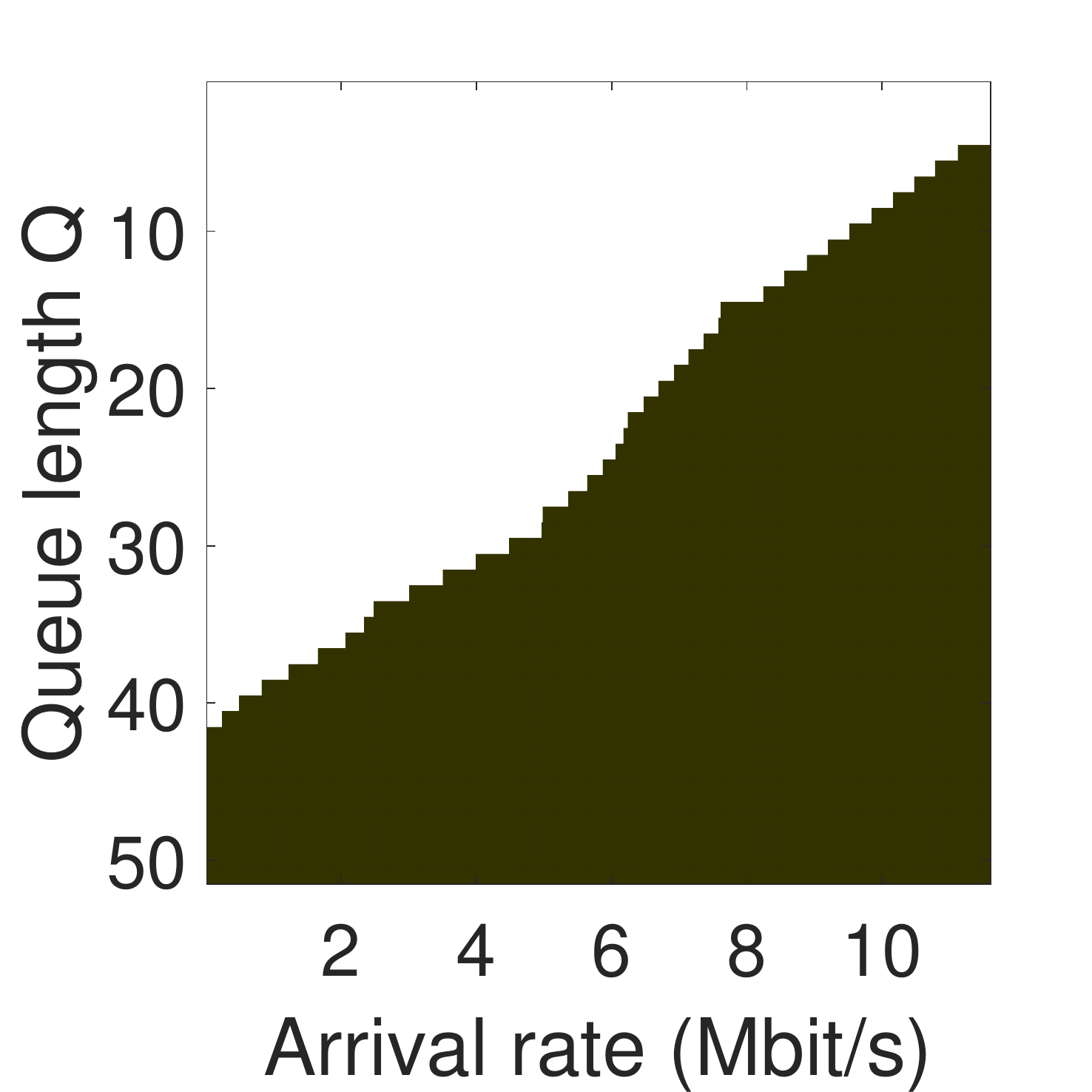}
		\caption{\label{fig:optPolicyHighRtt2}%
        RTT = $10$ ms}
	\end{subfigure}
\caption{\label{fig:optPolicyRtt}%
PAQMAN with known RTT: the left subplot assumes an RTT of $2$~ms and the right corresponds to a higher RTT of $10$~ms. Like Fig.~\ref{fig:optPolicyHighRtt2}, we take arrival rate $\in [0.01, 12]$~Mbit/s, service rate = $10$ Mbit/s, and a buffer size of $50$ packets and show the packet drop area in dark.
As PAQMAN admits all packets arriving between two decision events which is at least one RTT long, it compensates by dropping packets in (b) earlier than (a). 
In contrast to queue length based AQM heuristics these figures show that an estimate of the arrival rate is crucial to optimize the AQM performance. 
\vspace{-3pt}}%
\end{figure}

\begin{figure*}[t!]
	\centering
	\begin{subfigure}{0.32\textwidth}
		\centering
		\includegraphics[width=1\textwidth]{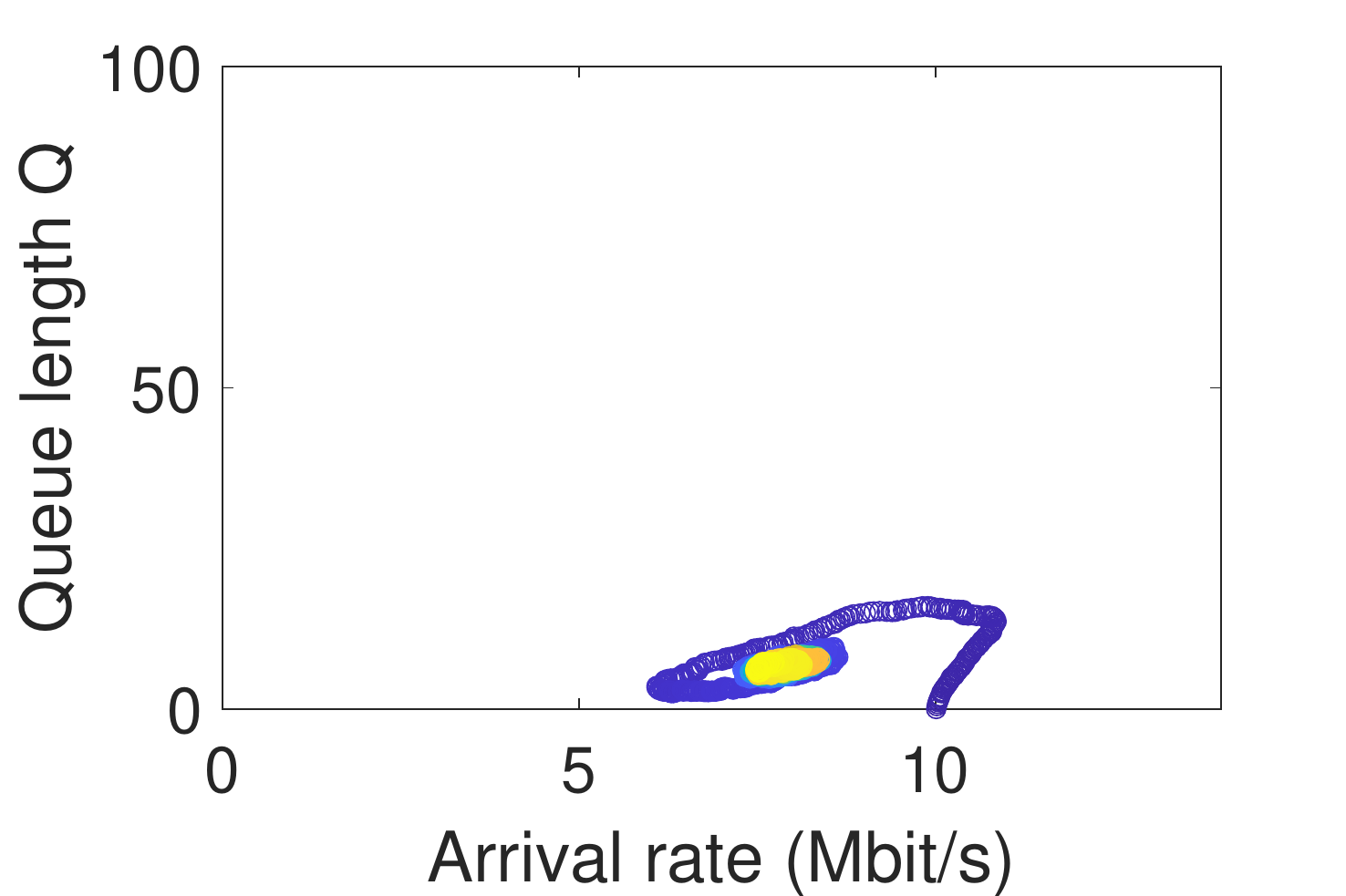}
		\caption{\label{fig:evolAqm10}%
		PAQMAN}
	\end{subfigure}
	    \hfill
	\begin{subfigure}{0.32\textwidth}
		\centering
		\includegraphics[width=1\textwidth]{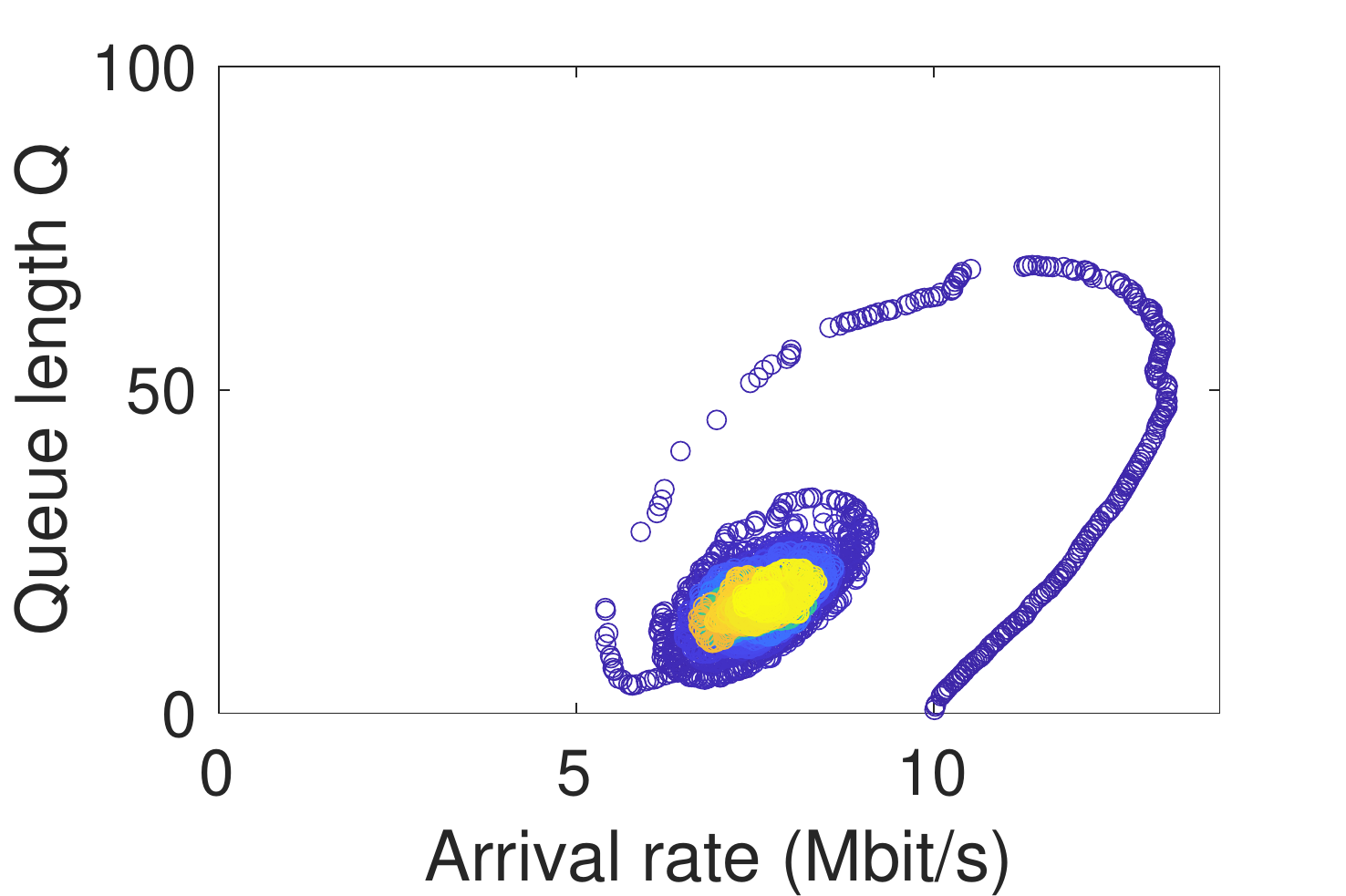}
		\caption{\label{fig:evolCodel10}%
		CoDel}
	\end{subfigure}
    \hfill
	\begin{subfigure}{0.32\textwidth}
		\centering
		\includegraphics[width=1\textwidth]{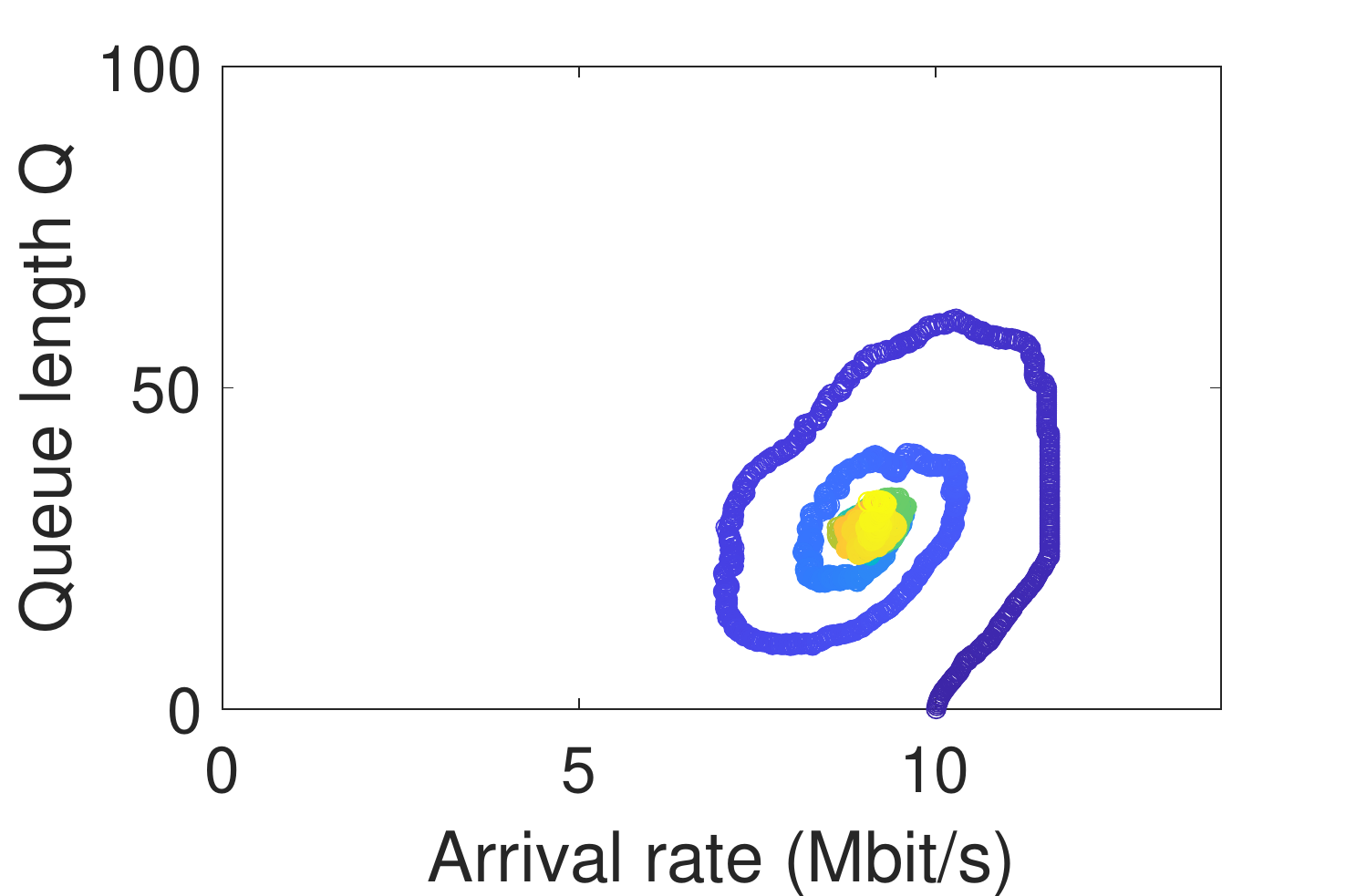}
		\caption{\label{fig:evolTailDrop10}%
        Droptail}
	\end{subfigure}%
\caption{\label{fig:evolCompareWS}%
State evolution under different policies for a switch port with  service rate~$=10$~Mbit/s. 
As already known, the droptail queue achieves higher stationary utilization than the AQMs at the cost of higher delay. PAQMAN leads to much shorter delay than CoDel, while achieving  comparable stationary throughput and faster convergence to steady-state.}%
\end{figure*}

In this section, we evaluate PAQMAN for TCP Reno traffic under different network scenarios using simulations.
We first focus on the case where the switch deals with a single flow having negligible RTT. The evaluation under non-negligible RTT is subsequently taken up in Sect.~\ref{subsec:evalRttSF}. Throughout our simulations, PAQMAN aims to achieve a delay shorter than the target delay $\eta = 50$~ms while optimizing the throughput. Recall from \eqref{eq:rewardFnTput} and \eqref{eq:rewardRtt} that the reward function combines individual delay and throughput objectives. The penalty amount in the reward function for breaching this delay is fixed at $M = 10^6$. To make comparisons fair, we set the target delay parameter of CoDel to the same target delay, wherever applicable. All other internal Codel parameters are left unchanged.  

While comparing the performance of different policies, we look at the empirical stationary behaviour of the corresponding system. 
Here, we consider the behavior of long-lived flows. To derive the stationary behaviour, we conduct $200$ simulation runs. Each run spans across $5 \times 10^4$ packet arrivals to simulate the stationary behaviour and the run starts with an arrival rate equal to the link service rate.
To generate comparison plots, we subsequently time-average the congestion indicators or the state of the system, given by the buffer filling of the switch and the flow arrival rate. Note that the stationary arrival rate equals the stationary throughput and the queue length determines the \emph{expected} delay. We use these immediate connotations to interpret the findings from our plots. To show the transient behaviour of the respective AQM algorithm in these plots, time propagation is indicated by varying the colour of the state from blue to yellow. \looseness = -1

\subsection{Negligible RTT} \label{subsec:zeroRtt}
In this section, we analyze PAQMAN for the case of a single flow having negligible RTT.
We follow the method described in Sect.~\ref{sec:model} to compute the packet drop policy. 

We first plot the derived policies using the inputs from~\eqref{eq:dataTransform} for two different average service rates.
The simulation setup for the first scenario corresponds to a service rate $\mu = 5$~Mbit/s, while the second is given by $\mu = 10$~Mbit/s.
Further, we calculate the policy for arrival rates in the range $[0.01, 12]$~Mbit/s, simulated using a Gamma distributed packet interarrival times with fixed shape parameter $\alpha = 1.5$. 
For our simulations, we only change the rate parameter $\beta$ of the Gamma random variable to reflect the impact of the drop/admit action on the effective arrival rate $\beta/\alpha$.
The resulting policies are shown in Fig.~\ref{fig:optPolicy}, where an incoming packet should be dropped if the current state of the switch, i.e., \textit{buffer filling and the packet arrival rate belong to the dark region}. As expected, a lower service rate at the switch entails more aggressive packet drops reflected by the difference in Fig.~\ref{fig:optPolicyLow} and \ref{fig:optPolicyHigh}. The non-trivial effect of flow arrival rate on the policy is also noteworthy. \looseness = -1

\begin{figure*}[t]
	\centering
	\begin{subfigure}{0.32\textwidth}
		\centering
		\includegraphics[width=1\textwidth]{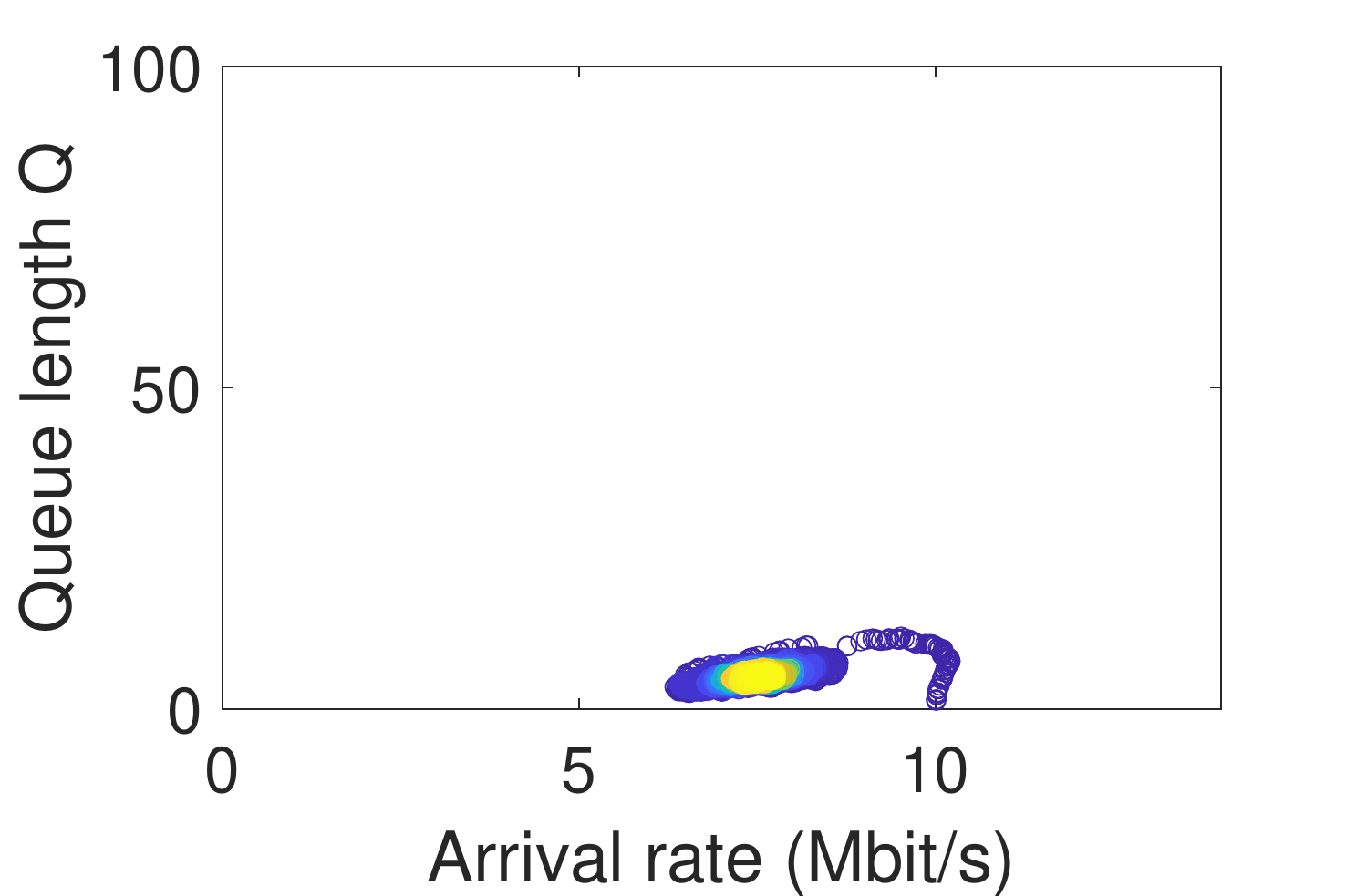}
		\caption{\label{fig:evolAqm10Rtt2}%
		PAQMAN}
	\end{subfigure}
	    \hfill
	\begin{subfigure}{0.32\textwidth}
		\centering
		\includegraphics[width=1\textwidth]{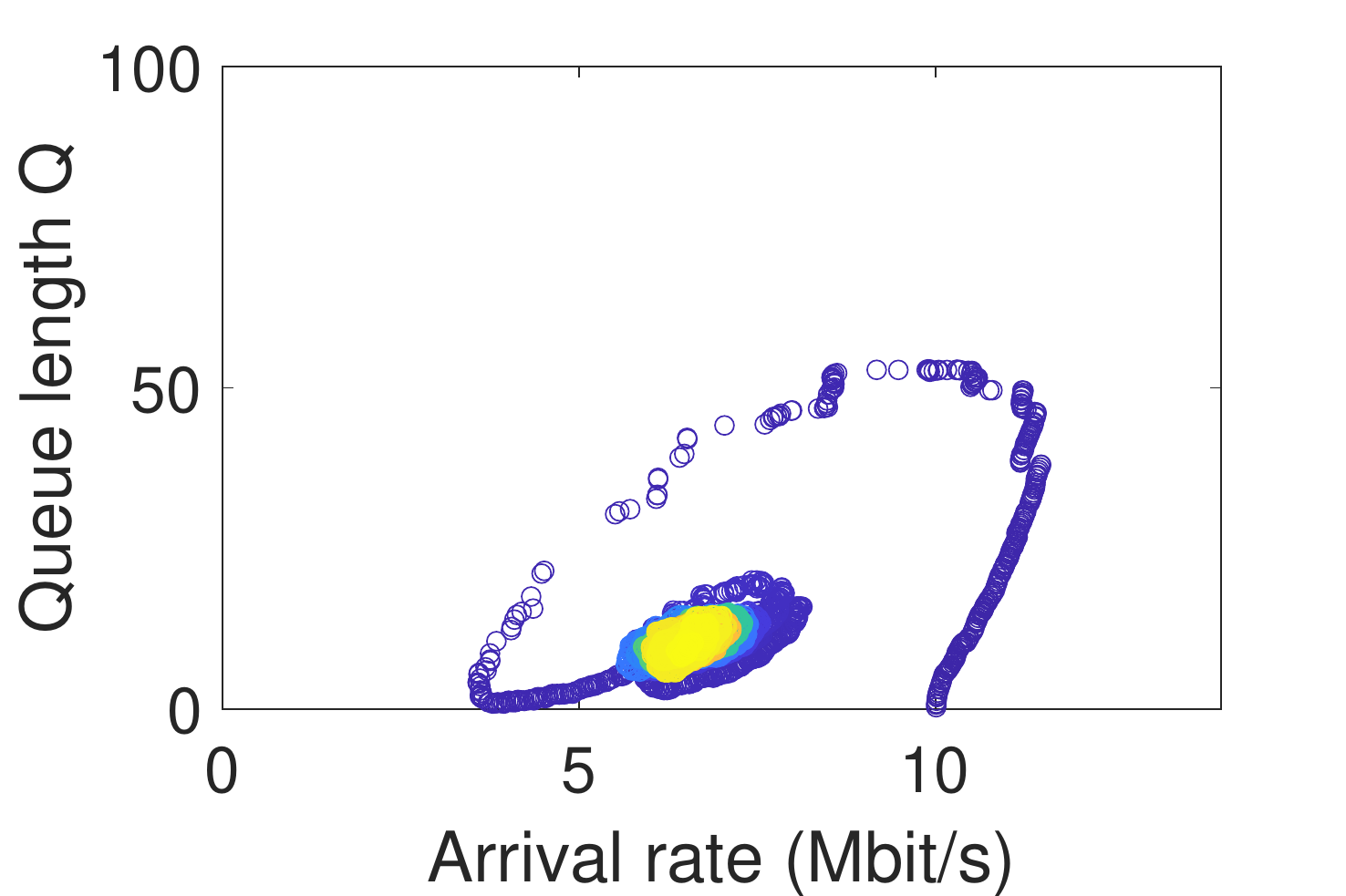}
		\caption{\label{fig:evolCodel10Rtt2}%
		CoDel}
	\end{subfigure}
    \hfill
	\begin{subfigure}{0.32\textwidth}
		\centering
		\includegraphics[width=1\textwidth]{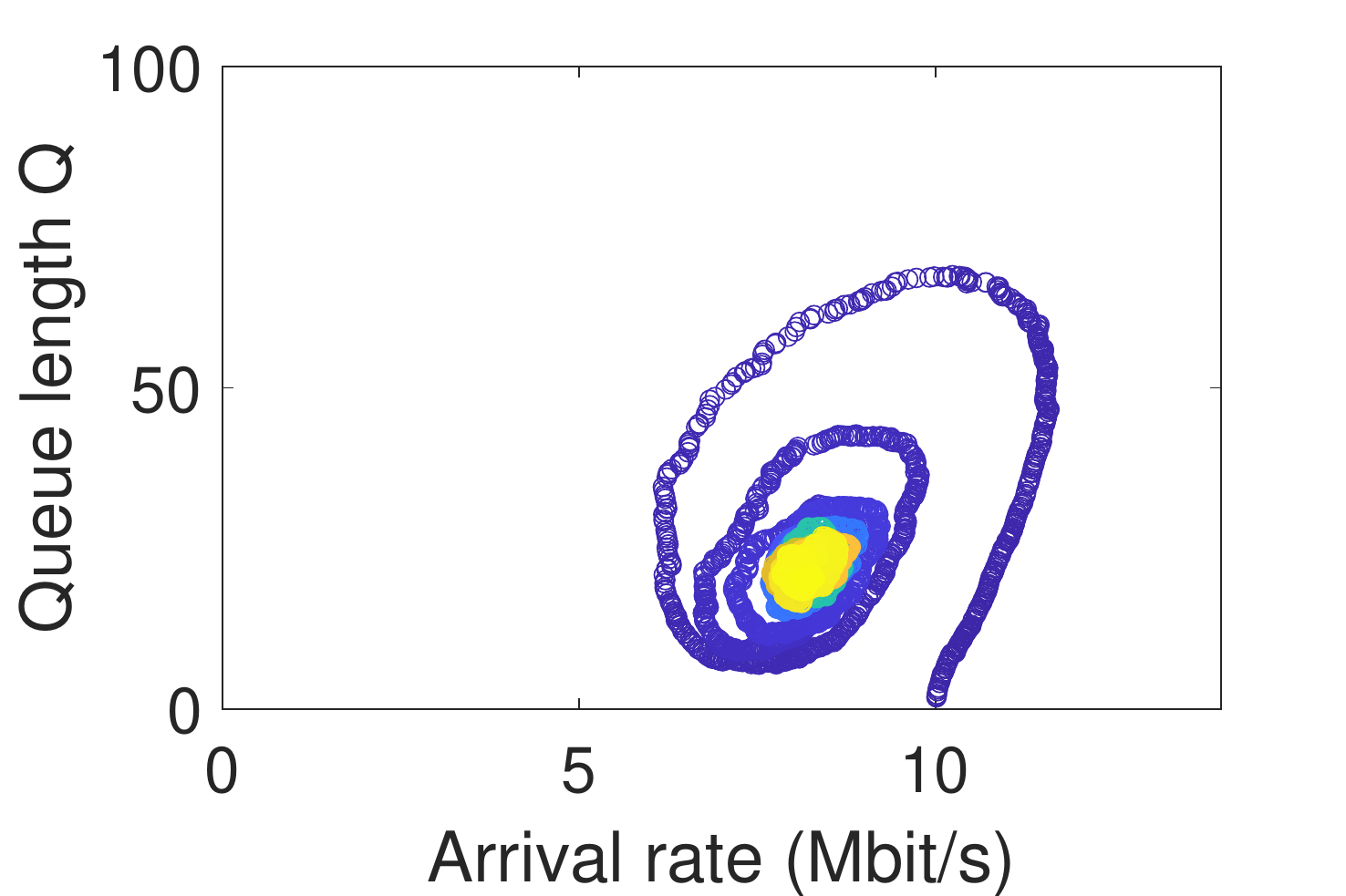}
		\caption{\label{fig:evolTailDrop10Rtt2}%
        Droptail}
	\end{subfigure}
\caption{\label{fig:evolCompare10msRtt2}%
State evolution under different policies for a flow with known RTT~$= 2$ ms and switch service rate~$= 10$~Mbit/s. Similar to Fig.~\ref{fig:evolCompareWS}, PAQMAN converges faster to the steady-state characterized by shorter delay and equivalent throughput.}%
\end{figure*}

\begin{figure*}[t!]
	\centering
	\begin{subfigure}{0.32\textwidth}
		\centering
		\includegraphics[width=1\textwidth]{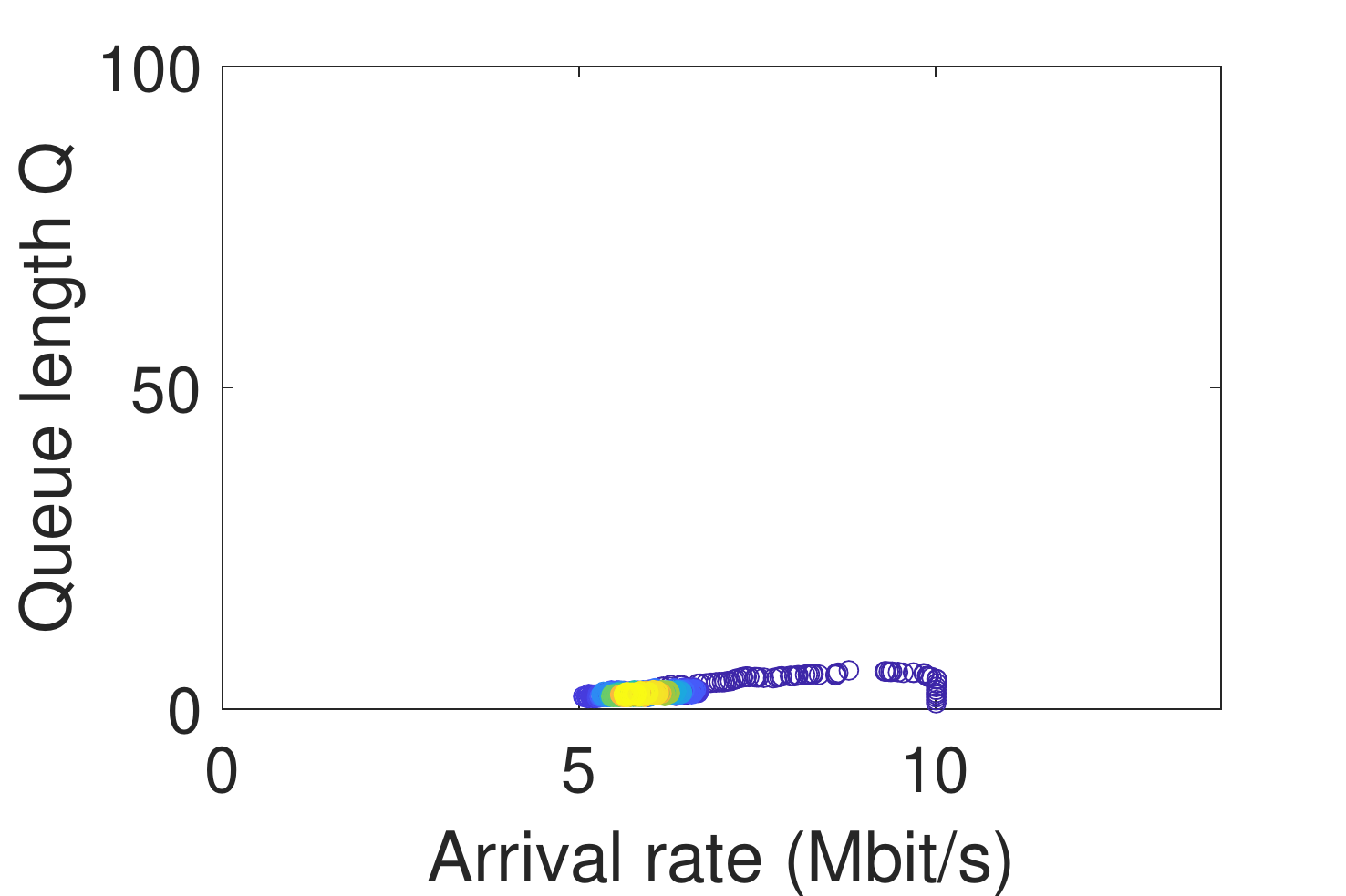}
		\caption{\label{fig:evolAqm10Rtt10.1}%
		PAQMAN}
	\end{subfigure}
	    \hfill
	\begin{subfigure}{0.32\textwidth}
		\centering
		\includegraphics[width=1\textwidth]{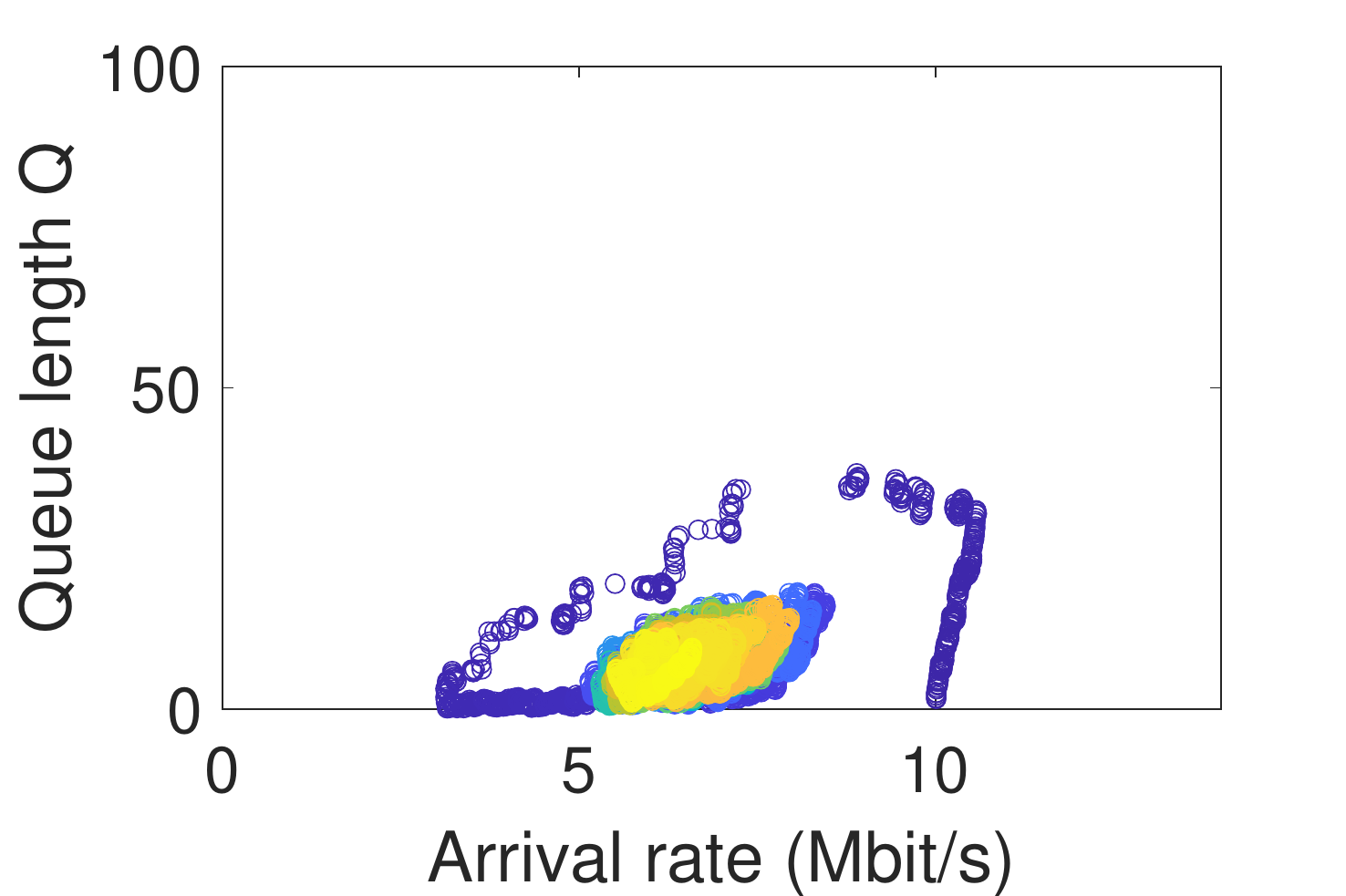}
		\caption{\label{fig:evolCodel10Rtt10.1}%
		CoDel}
	\end{subfigure}
    \hfill
	\begin{subfigure}{0.32\textwidth}
		\centering
		\includegraphics[width=1\textwidth]{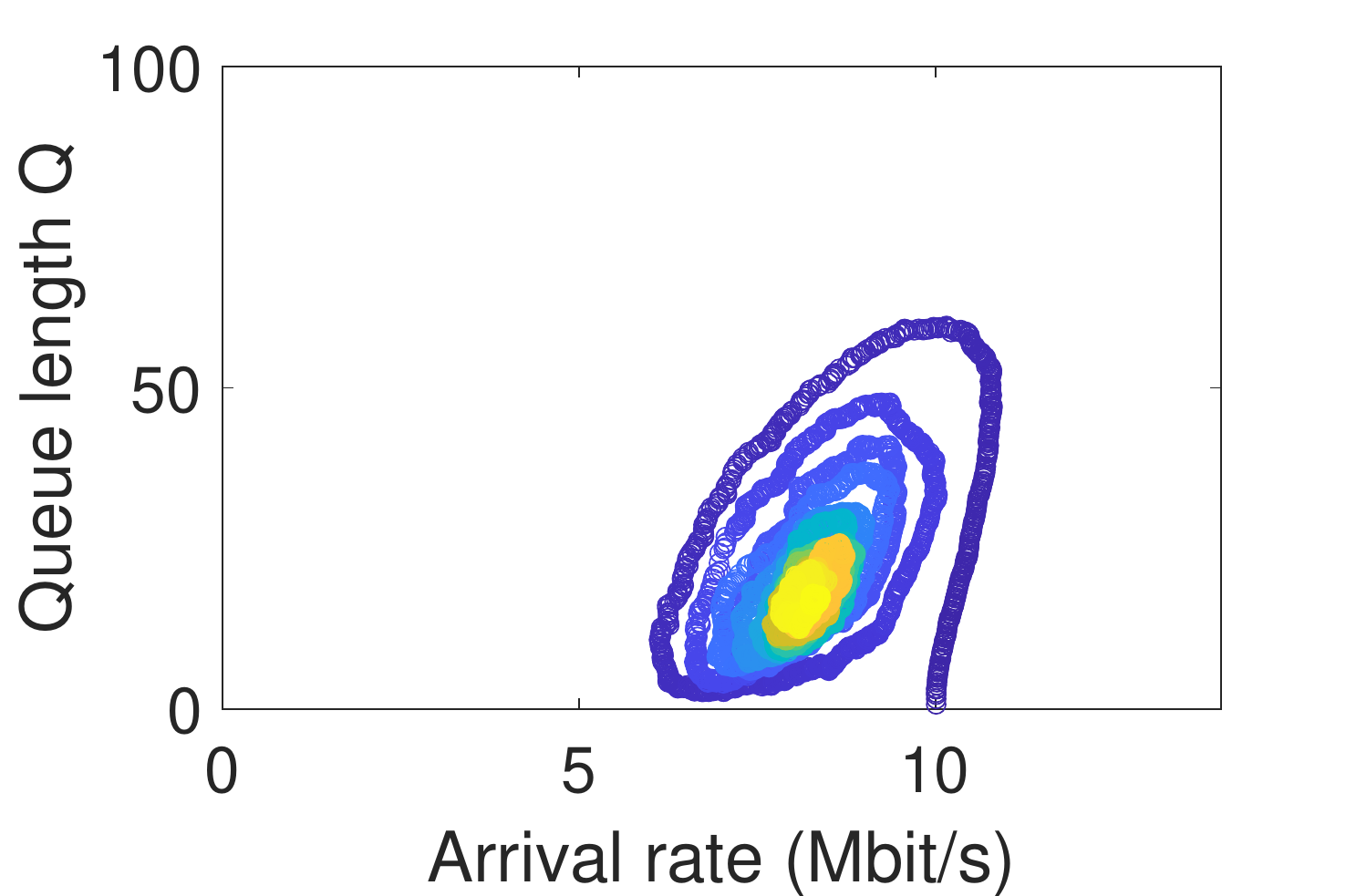}
		\caption{\label{fig:evolTailDrop10Rtt10.1}%
        Droptail}
	\end{subfigure}
\caption{\label{fig:evolCompare10msRtt10.1}%
State evolution under different policies for a setup same as Fig.~\ref{fig:evolCompare10msRtt2} except that the flow RTT is much higher ($10$ ms). We observe similar phenomena as Fig.~\ref{fig:evolCompare10msRtt2} although both the stationary delay and the throughput decrease across AQM policies.}%
\end{figure*}

Next, we compare PAQMAN to CoDel and droptail queues. As mentioned earlier, the target delay parameter of CoDel is set to the delay threshold for PAQMAN. The evaluation is shown in Fig.~\ref{fig:evolCompareWS} where the policy from Fig.~\ref{fig:optPolicyHigh} is used to generate the left subplot.
We plot the system state, given by the \textit{(i)} current buffer filling and \textit{(ii)} flow arrival rate, time-averaged over multiple runs. 
Our plots suggest that, in stationarity, PAQMAN results in an arrival rate that is comparable to CoDel, although the queue length appears to be much shorter. This immediately translates to the fact that PAQMAN yields equivalent stationary throughput while keeping the delay much shorter. As already known, the droptail policy in Fig.~\ref{fig:evolTailDrop10} generates near-perfect utilization at the cost of a longer delays.

\subsection{Non-negligible RTT} \label{subsec:evalRttSF} 

Next, we consider the case when the flow RTT is non-negligible. We assume throughout this subsection that the RTT is estimated by (\emph{or} known to) the switch. We follow the method from Sect.~\ref{sec:rttModel} to derive PAQMAN under each setting and accordingly simulate the system. As before, we compare the resulting performance with CoDel and droptail queues. \looseness = -1 

\begin{figure*}[t]
	\centering
	\begin{subfigure}{0.32\textwidth}
		\centering
		\includegraphics[width=1\textwidth]{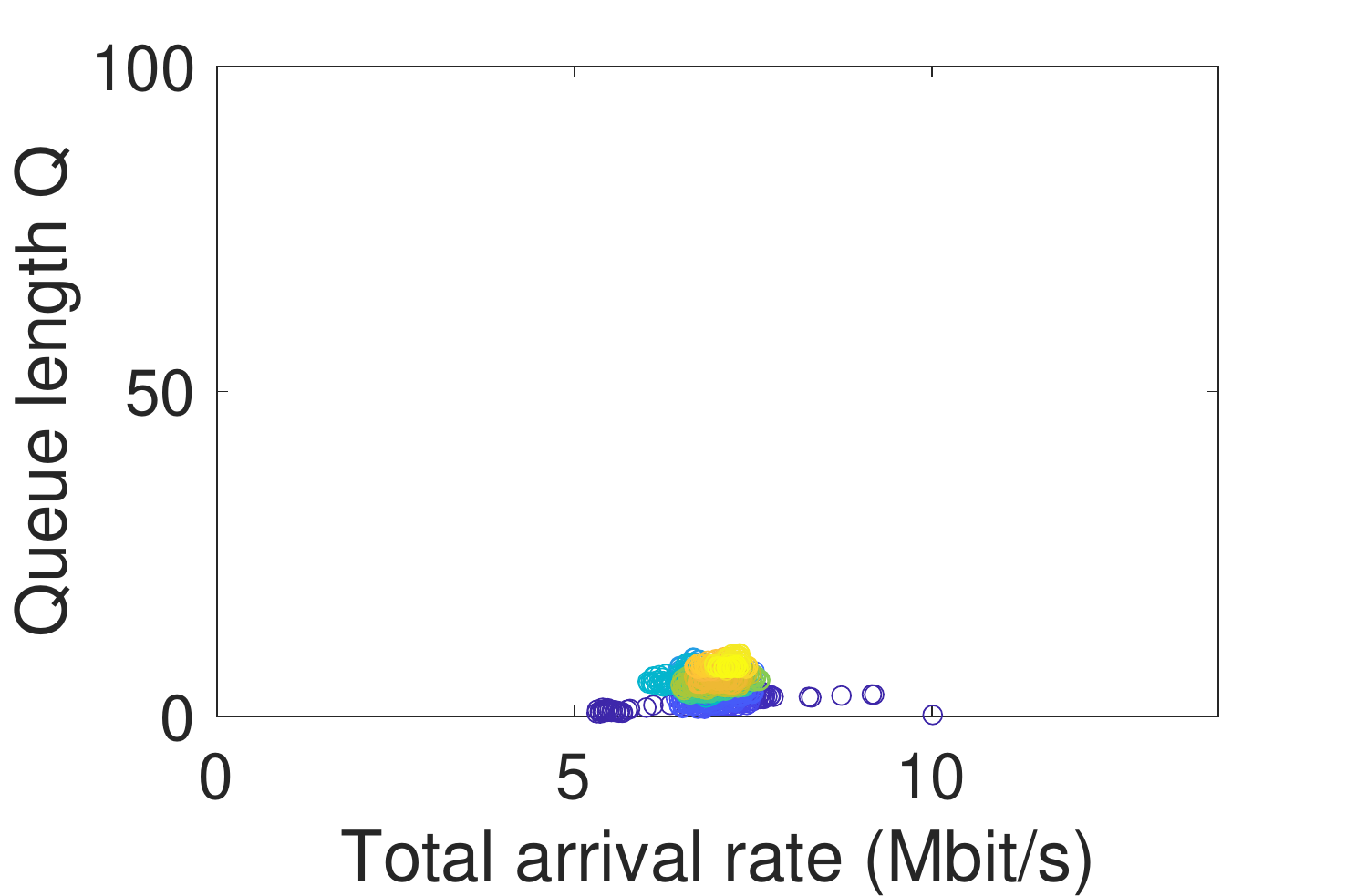}
		\caption{\label{fig:evolAqmMF246}%
		PAQMAN}
	\end{subfigure}
	    \hfill
	\begin{subfigure}{0.32\textwidth}
		\centering
		\includegraphics[width=1\textwidth]{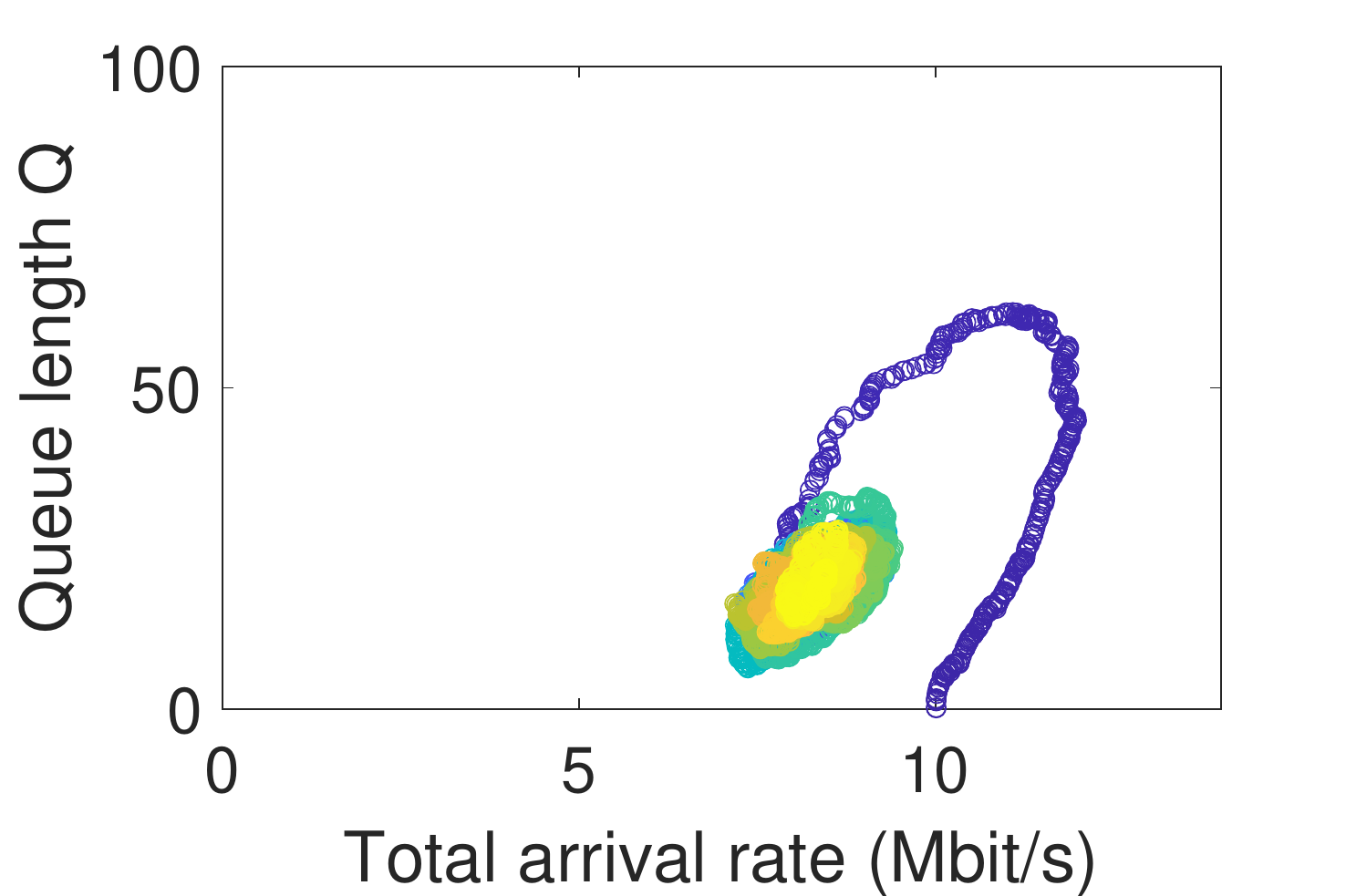}
		\caption{\label{fig:evolCodelMF246}%
		CoDel}
	\end{subfigure}
    \hfill
	\begin{subfigure}{0.32\textwidth}
		\centering
		\includegraphics[width=1\textwidth]{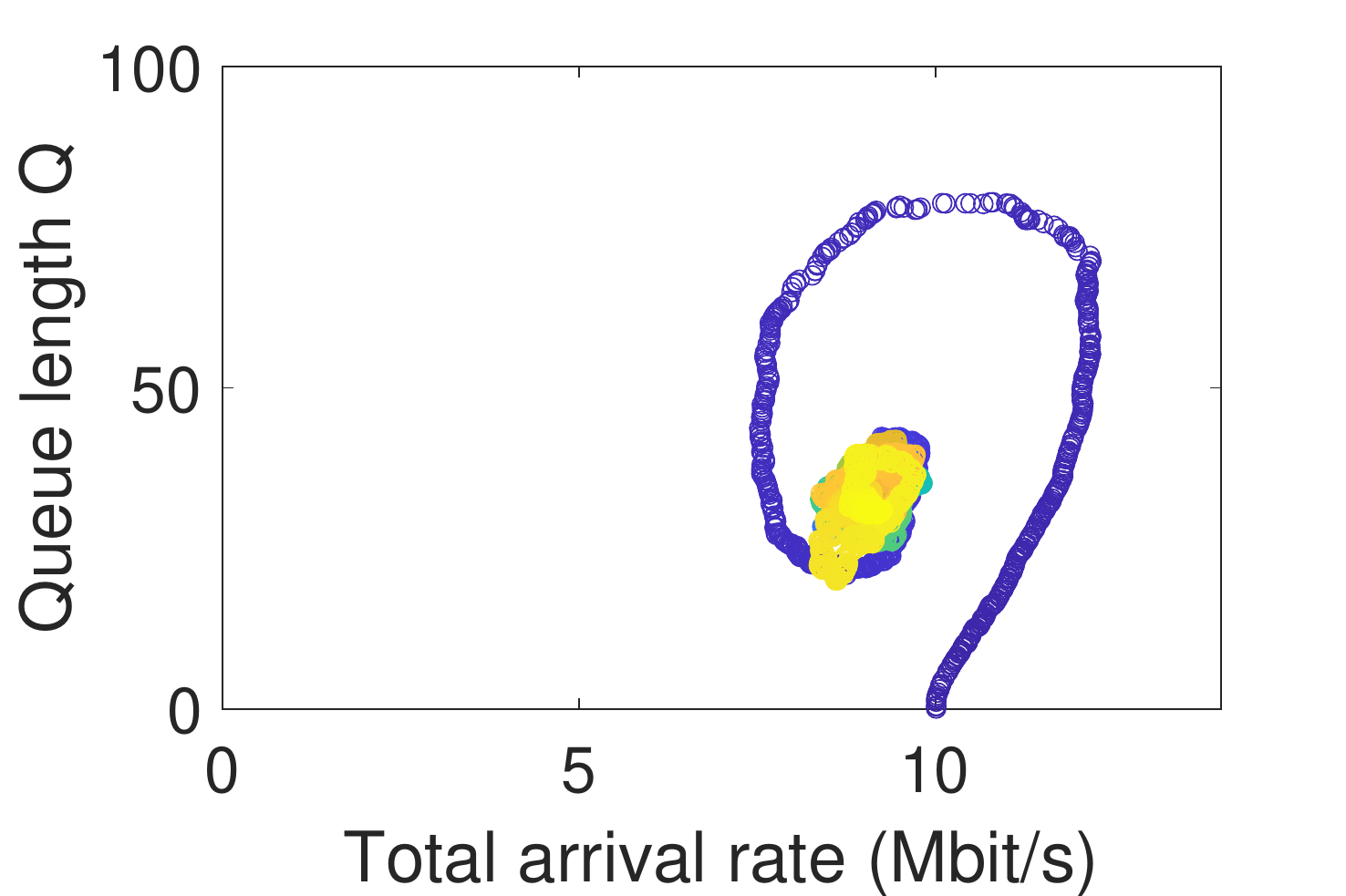}
		\caption{\label{fig:evolTailDropMF246}%
        Droptail}
	\end{subfigure}%
\caption{\label{fig:evolCompare10mbRtt246}%
Evolution of queue length and the total (aggregate) arrival rate for $3$ flows under different policies for service rate = $10$ Mbit/s with target delay $= 50$ ms and RTT = $= [2, 4, 6]$ ms.}%
\end{figure*}

In Fig.~\ref{fig:optPolicyRtt}, we illustrate PAQMAN under low and high RTTs. We fix the packet service rate at $10$~Mbit/s and restrict our attention to arrival rate $\in [0.01, 12]$~Mbit/s. 
Recall that in this case, PAQMAN admitted all arrivals between two decision epochs subject to sufficient room at the buffer of the switch. This leads to increased packet buffering for higher RTTs, especially in high arrival regimes as the inter-decision times are RTT-based. To alleviate this problem, the policy is more aggressive and starts dropping earlier as seen via a comparison of Fig.~\ref{fig:optPolicyLowRtt2} and \ref{fig:optPolicyHighRtt2}.

\begin{figure}[t!]
	\centering
	\begin{subfigure}[t]{0.48\columnwidth}
		\centering
		\includegraphics[width=1\textwidth]{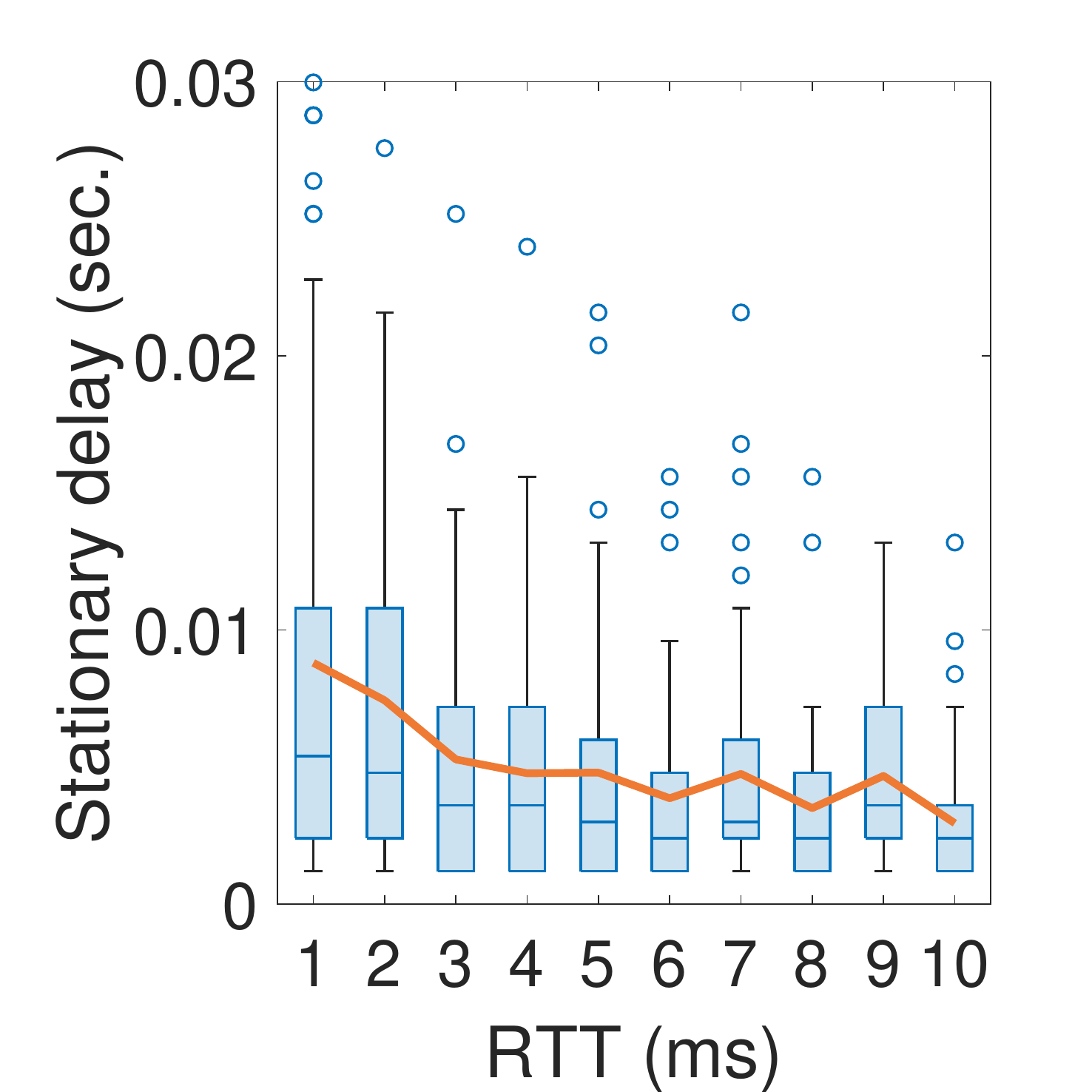}
		\caption{\label{fig:delaybyRTT}%
		Delay}
	\end{subfigure}
    \hfill
	\begin{subfigure}[t]{0.48\columnwidth}
		\centering
		\includegraphics[width=1\textwidth]{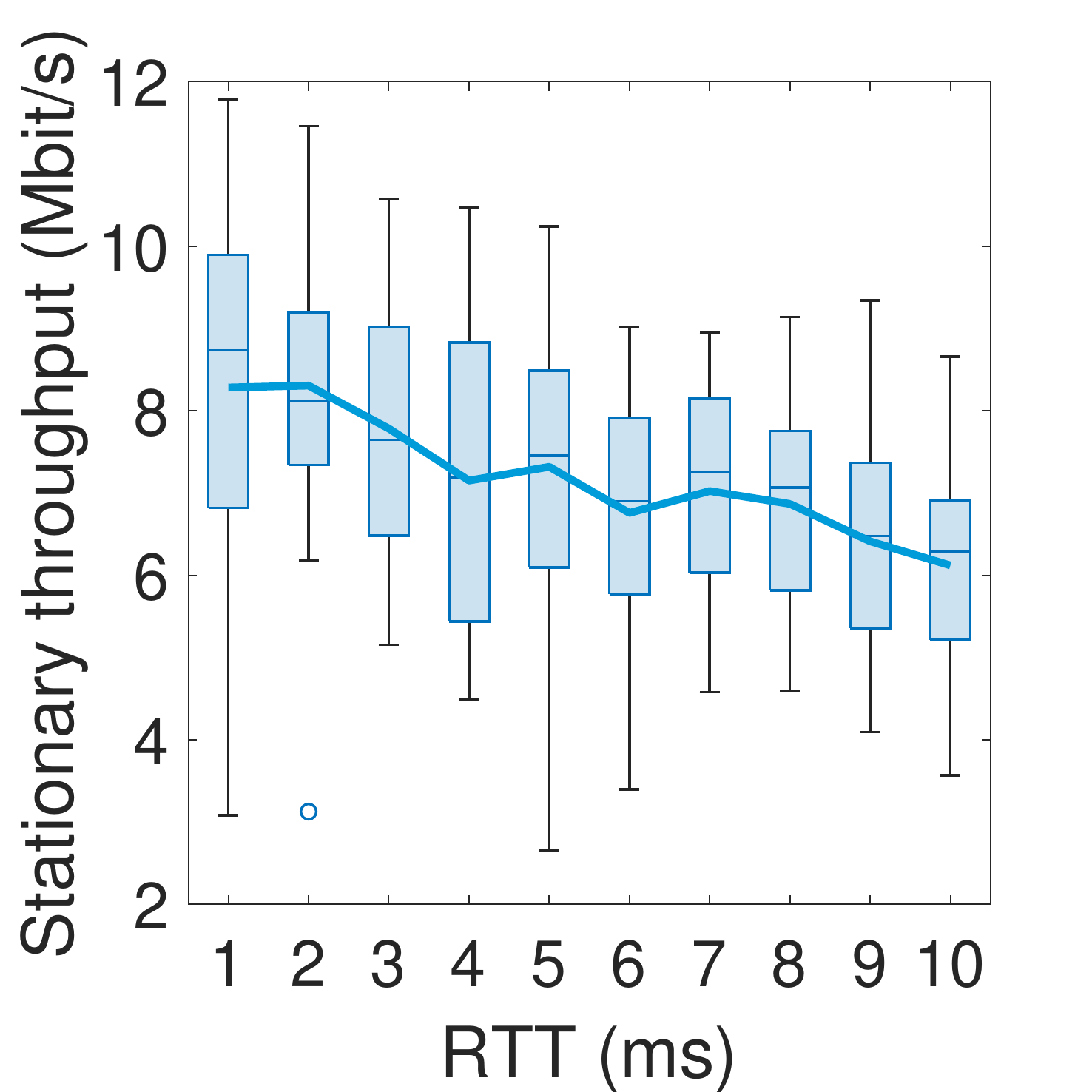}
		\caption{\label{fig:tPutbyRTT}%
        Throughput}
	\end{subfigure}
\caption{\label{fig:byRTT}%
Impact of flow RTT on the stationary delay and throughput. Given an RTT, we simulate the system for long-lived flows using PAQMAN. Each box plot is generated using $200$ simulation runs. Consistent with previous plots, both delay and throughput decrease with an increasing RTT.}%
\end{figure}

In Fig.~\ref{fig:evolCompare10msRtt2} and \ref{fig:evolCompare10msRtt10.1}, we compare PAQMAN with CoDel and Droptail when there is only one flow passing through the buffered switch port. The state evolution under PAQMAN (left subplot) in these figures is generated using the corresponding policy from Fig.~\ref{fig:optPolicyRtt}, which follows the workflow described in Sect.~\ref{sec:rttModelSF}. Similar to Fig.~\ref{fig:evolCompareWS}, we see that PAQMAN achieves shorter delay than CoDel while yielding comparable stationary throughput. Further, a quick comparison of Fig.~\ref{fig:evolCompareWS}-\ref{fig:evolCompare10msRtt10.1} reveals that a longer flow RTT leads to a reduction in both throughput and delay, which we investigate further in Fig.~\ref{fig:byRTT}. 
Here, we first derive PAQMAN and subsequently simulate the system for long-lived flows ($5 \times 10^4$ packet arrivals) for every given RTT. The box plots per RTT in Fig.~\ref{fig:byRTT} are generated using $200$ simulation runs. Consistent with Fig.~\ref{fig:evolCompareWS}-\ref{fig:evolCompare10msRtt10.1}, we see that the stationary delay and the stationary throughput tend to diminish as the RTT increases. 
\vspace{-1.5pt}

Finally, in Fig.~\ref{fig:evolCompare10mbRtt246}, we compare PAQMAN's performance in the multi-flow case. 
Here, we consider three flows with RTT $[2, 4, 6]$ ms and the service is assumed to be $\mu = 10$ Mbit/s. 
We start the system in a phase where the total arrival rate equals the switch port service rate and the individual arrival rates are equal, i.e., each flow has equal share of the bandwidth. 
Similar to the experiments in the single flow case, we simulate the system for $200$ runs, where each run spans across $5 \times 10^4$ packet arrivals. 
To learn the policy, we pre-simulate the system and use a DQN to approximate the Q-values for each system state, which in turn decides the optimal action. 
As with any offline-learned  algorithm, the efficacy can be verified only under the assumption that underlying conditions do not vary much from the training sample which we ensure during the simulation.
The time-averaged plot of the aggregated system state expressed in terms of queue length and total arrival rate is shown in Fig.~\ref{fig:evolCompare10mbRtt246}. 
We see that that PAQMAN achieves similar throughput to CoDel, while keeping the delay considerably low here as well.

\section{Conclusion}\label{sec:conclusion}

In this paper we presented PAQMAN, a principled approach to AQM with corresponding algorithmic design. 
In contrast to many AQM heuristics, we take a direct approach to incorporate the packet arrival rate in our formulation of the state of the system and optimize with respect to a linear combination of delay and throughput objectives.
Our work takes a probabilistic approach to combine a model of congestion control (mainly Additive Increase- Multiplicative Decrease) with AQM packet drops to formulate the AQM problem as finding optimal packet dropping policy in a Semi-Markov Decision Process, given a target delay parameter. 
Simulation results show that the direct approach of incorporating the arrival rate in the state leads to comparable throughput of the system to the widely used AQM policy CoDel, while outperforming it in terms of latency.
\looseness=-1

\section{Appendix}
\label{sec:appendix}

\begin{proof}[Proof of Lemma~\ref{lemma:gammaRec}]:
Let us first denote the upper incomplete gamma integral by $\gamma(\cdot,\cdot)$, which satisfies the following recursion:
\begin{align*}
    \gamma(x+1,y) = \int_y^{\infty} t^{x} e^{-t}dt
                  = x \gamma(x,y) + y^x e^{-y},
\end{align*}
for $\Re(x)>0$. We see that
\begin{align*}
    & \quad \mathsf{P}(Y_{u,v}>X_{w,z})\\
    &= \int_{0}^{\infty}\mathsf{P}(Y_{u,v}>X_{w,z}|X_{w,z}=t)dP_{X}(t)\\
    &= \int_{0}^{\infty}\mathsf{P}(Y_{u,v}>t)dP_{X}(t), \; Y_{u,v} \indep X_{w,z}\\
    &= \int_{0}^{\infty}\frac{\gamma(u,vt)}{\Gamma (u)}\frac{z^w t^ {w-1} e^{-z t}}{\Gamma (w)} dt\\
    &= \int_{0}^{\infty}\frac{(u-1)\gamma(u-1,vt)+(vt)^{u-1} e^{-vt}}{\Gamma (u)}\frac{z^w t^ {w-1} e^{-z t}}{\Gamma (w)} dt\\
    &= \mathsf{P}(Y_{u-1,v}>X_{w,z}) + \int_{0}^{\infty}\frac{v ^ {u-1}z^w t^ {u+w-2} e^{-(v+z) t}}{\Gamma (u)\Gamma (w)} dt\\
    &= \mathsf{P}(Y_{u-1,v}>X_{w,z}) + \frac{\Gamma (u+w-1) v ^ {u-1}z^w}{\Gamma (u)\Gamma (w)(v+z)^{u+w-1}} \cdot \\
    & \quad \int_{0}^{\infty}\frac{(v+z)^{u+w-1} t^ {u+w-2} e^{-(v+z) t}}{\Gamma (u+w-1)} dt\\
    &= \mathsf{P}(Y_{u-1,v}>X_{w,z}) + \\
    & \quad \frac{\Gamma (u+w-1)}{\Gamma (u)\Gamma (w)} \bigg(\frac{v}{v+z}\bigg) ^ {u-1} \bigg(\frac{z}{v+z}\bigg) ^ w,
\end{align*}
which proves the first part of the lemma. 
By induction on $u$,
\begin{align}\label{gammaRec}
\begin{aligned}
    & \mathsf{P}(Y_{u,v}>X_{w,z}) = \mathsf{P}(Y_{1,v}>X_{w,z}) \; + \\
    & \quad \sum_{k=2}^{u} \frac{\Gamma (k+w-1)}{\Gamma (k)\Gamma (w)} \bigg(\frac{v}{v+z}\bigg) ^ {k-1} \bigg(\frac{z}{v+z}\bigg) ^ w~.
\end{aligned}
\end{align}
Since $Y_{1,v}$ is exponentially distributed with parameter $v$,
\begin{align*}
    \mathsf{P}(Y_{1,v}>X_{w,z}) 
    &= \int_{0}^{\infty}e^{-v t}\frac{z^w t^ {w-1} e^{-z t}}{\Gamma (w)} dt \nonumber \\
    &= \bigg(\frac{z}{v+z}\bigg) ^ w \int_{0}^{\infty}\frac{(v+z)^w t^ {w-1} e^{-(v+z) t}}{\Gamma (w)} dt \nonumber \\
    &= \bigg(\frac{z}{v+z}\bigg) ^ w~.
\end{align*}
Hence from \eqref{gammaRec}, 
\begin{align*}
    \mathsf{P}(Y_{u,v}>X_{w,z})= \sum_{k=0}^{u-1} \frac{\Gamma (k+w)}{\Gamma (k+1)\Gamma (w)} \bigg(\frac{v}{v+z}\bigg) ^ k \bigg(\frac{z}{v+z}\bigg) ^ w~.
\end{align*}
\end{proof}

\begin{proof}[Proof of Prop.~\ref{coro:transProbSmdp}]:
We start by observing that $\mathsf{P}((q,\beta_{t+1})|S_t,0)$ denotes the probability of the event that exactly $Q_t+1-q$ packets are served between two packet arrivals for $1 \le q \le Q_t+1$. Now, at least $n$ packets are served between two arrivals 
if and only if the corresponding service time $Y_{n,\mu}$ is less than the respective interarrival time $X_{\alpha,\beta_{t+1}}$. The change in the arrival rate parameter is given by the fact that following an action on the $t$-th packet arrival, the arrival rate parameter is instantly changed to $\beta_{t+1} = \beta_t+\alpha$ which \textit{determines the distribution of the next interarrival time}. Thus, given $\beta_{t+1} = \beta_t+\alpha$, for $1 \le q \le Q_t+1$,
\begin{align*}
    & \quad \thickspace \mathsf{P}((q,\beta_{t+1})|S_t,0)\\
    &= \mathsf{P}(Y_{Q_t+2-q,\mu}>X_{\alpha,\beta_{t+1}})-\mathsf{P}(Y_{Q_t+1-q,\mu}>X_{\alpha,\beta_{t+1}})\\
    &=  \frac{\Gamma (Q_t+1-q+\alpha)}{\Gamma (Q_t+2-q)\Gamma (\alpha)} \bigg(\frac{\mu}{\mu+\beta_{t+1}}\bigg) ^ {Q_t+1-q} \bigg(\frac{\beta_{t+1}}{\mu+\beta_{t+1}}\bigg) ^ {\alpha}~.
\end{align*}
For $q=0$, we observe that no more than $Q_t+1$ packets can be served between two arrivals.
Hence, given $\beta_{t+1} = \beta_t+\alpha$, 
\begin{align*}
    & \quad \mathsf{P}((0,\beta_{t+1})|S_t,0)\\
    &= \mathsf{P}(Y_{Q_t+1,\mu} \le X_{\alpha,\beta_{t+1}}) \\
    &= 1 - \mathsf{P}(Y_{Q_t+1,\mu} > X_{\alpha,\beta_{t+1}}) \\
    &= 1 - \sum_{k=0}^{Q_t} \frac{\Gamma (k+\alpha)}{\Gamma (k+1)\Gamma (\alpha)} \bigg(\frac{\mu}{\mu+\beta_{t+1}}\bigg) ^ {Q_t+1-k} \bigg(\frac{\beta_{t+1}}{\mu+\beta_{t+1}}\bigg) ^ {\alpha}~,
\end{align*}
which completes the proof.
\end{proof}

\begin{proof}[Proof of Prop.~\ref{thm:transitionProbRtt}]
The queue length grows like the population of a birth-death process in the interval $(T_i, T_i+r]$ with intensity matrix $G_i$. Thus the transition matrix for this interval is given by $e^{r G_i}$.
Further $T_{i+1}-(T_i+r) \sim \text{Exp}(\beta_i)$ implying
\begin{align*}
    P_i 
    = e^{r G_i} \int_0^\infty e^{t G} \beta_i e^{-\beta_it} dt, 
    = \beta_i e^{r G_i} \int_0^\infty e^{-(\beta_i I -G) t} dt.
\end{align*}
We have used the facts that 
$e^{-c}I = e^{-cI}$ and $cI$ commutes with any matrix of similar dimension $\forall c \in \setOfReals$. Now,
{\small $$ \beta_i I -G = 
\begin{bmatrix}
\beta_i & 0 & 0 & \dots & \dots\\
-\mu & \mu+\beta_i & 0 & 0 & \dots\\
0 & -\mu & \mu+\beta_i & 0 & \dots\\
\dots & \dots & \dots & \dots & \dots\\
0 & \dots & \dots & -\mu & \mu+\beta_i
\end{bmatrix}.$$
}
Therefore, by Gershgorin's theorem~\cite{zbMATH02560682}, all eigenvalues of $\beta_i I -G$ lie in the following closed balls: $B(\beta_i, 0), 
B(\mu+\beta_i, \mu)$. Hence, by lemma \ref{lemma:expmIntegral}
$$
P_i = \beta_i e^{G_i r} (\beta_i I -G)^{-1}.
$$
\end{proof}

\begin{proof}[Proof of~\eqref{eq:transKernelFactorization}]:
Observe that
\begin{align} \label{eq:conditionalMultRtt}
\begin{aligned}
    & P \bigg(Y_t \in (a,b], Y_t = Y_{jt}, V_t = Q_{t+1}-Q_t \bigg)  \\
    = \thinspace & \int_a^b P \bigg(Y_t = y, V_t = Q_{t+1}-Q_t|Y_{jt} = y\bigg) dF_{Y_{jt}}(y)   \\
    = \thinspace & \mathbbm{1}_{j\in \{i_1 \dots i_l\}} \int_a^b P \bigg(\bigcap\limits_{k \le l, i_k \ne j} Y_{i_k t} > y\bigg)  \\
    &  \mathsf{P}(V_t = Q_{t+1}-Q_t|Y_{t} = y) dF_{Y_{jt}}(y) \\
    = \thinspace & \mathbbm{1}_{j\in \{i_1 \dots i_l\}}  \int_a^b \prod\limits_{k \le j, i_k \ne j} \mathsf{P}(Y_{i_k t} > y) \\
    &  \mathsf{P}(V_t = Q_{t+1}-Q_t|Y_{t} = y) dF_{Y_{jt}}(y)~.
\end{aligned}
\end{align}
Here, the indicator function $\mathbbm{1}_{j\in \{i_1 \dots i_l\}}$ indicates the fact that in order to have the flow index of the packet involving the next decision epoch as $j$, $c_j \le c_{i_l}$ should hold. Further, the first term in the integrand corresponds to the fact that given $Y_{jt} = y$, the events $\{Y_t = Y_{jt}\}$ and $\{Y_{i_k t} > y~ \forall k \le l, i_k \ne j\}$ are equivalent. 
Now, to evaluate the expression $\mathsf{P}(V_t = Q_{t+1}-Q_t|Y_{t} = y)$ in \eqref{eq:conditionalMultRtt}, we notice that the queue length of the underlying process grows like the population of a birth-death process where the death rate is always $\mu$ subject to a positive queue length. The birth rate, however, depends upon the time interval. Observe that there cannot be any packet arrival from $i_k$-th flow once the time since the last decision epoch exceeds $c_{i_k}$. Thus, the $(L\!+\!1)\!\times\!(L\!+\!1)$ intensity matrix in the interval $(c_{i_{k-1}},c_{i_k}], 1 \le k \le n$ is given by
$$
H_{i_k} = 
\begin{bmatrix}
-b_k & b_k & 0 & \dots & \dots\\
\mu & -\mu-b_k & b_k & 0 & \dots\\
0 & \mu & -\mu-b_k & b_k & \dots\\
\dots & \dots & \dots & \dots & \dots\\
0 & \dots & \dots & \mu & -\mu
\end{bmatrix}~, 
$$
with the sum of flow rates $b_k = \sum_{m \ge k} \beta_{i_m}$ and $c_{i_0}=0$ by definition. Further, the intensity matrix in the interval $(c_{i_n},\infty)$ is given by $G$ from \eqref{eq:G0}. The different intensity matrices for the arrivals in the different time spans after the last AQM decision is illustrated in Fig. \ref{fig:kernelQLenPartSplit}.

\begin{figure}[t]
	\centering
	\includegraphics[width=0.95\linewidth]{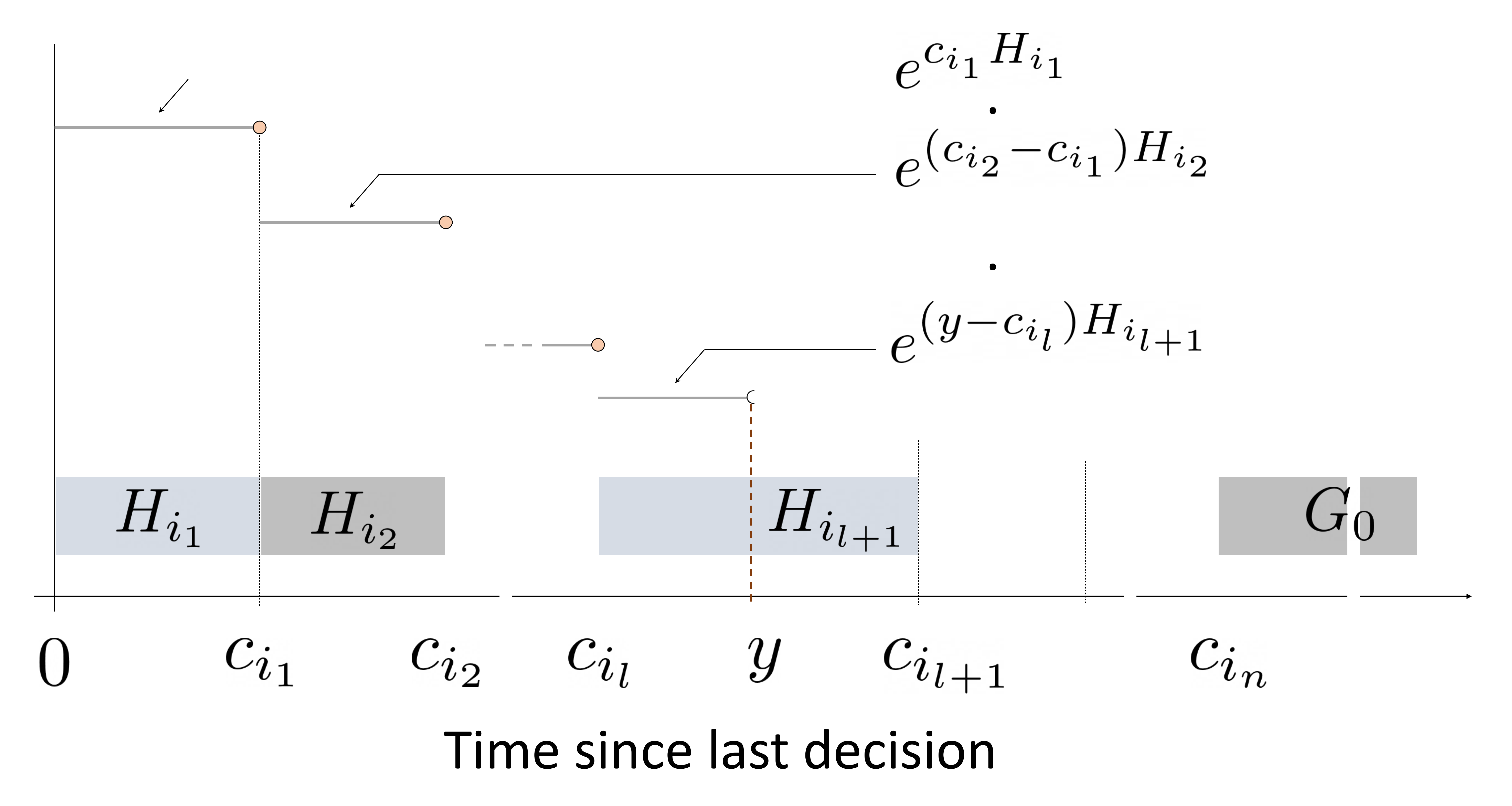}
    \caption{Intensity and corresponding transition matrices (TMs) between two decision epochs: for the time since last decision epoch, the intensity matrices ($H$) are highlighted in grey for successive intervals. Corresponding TMs over a (sub) interval of length $x$ is given by $e^{xH}$. The TM for an interval consisting of contiguous subintervals is the product of respective transition matrices as shown for $(0,y)$. 
    }
    \label{fig:kernelQLenPartSplit}
\end{figure}

Replacing $\mathsf{P}(V_t = Q_{t+1}-Q_t|Y_{t} = y)$ in \eqref{eq:conditionalMultRtt} by the corresponding transition matrix and denoting the respective transition kernel for $P (Y_t \in (a,b], Y_t = Y_{jt}, V_t = Q_{t+1}-Q_t)$ as $G_j(a,b)$, we obtain the formulation~\eqref{eq:transKernelFactorization}.

\end{proof}
%
\balance	
\bibliographystyle{IEEEtran}
\bibliography{sample-bibliography}

\end{document}